\documentclass[usenatbib]{mn2e}
\usepackage{verbatim}
\usepackage{graphicx}
\usepackage{url}
\begin{document}
\title{Human Contrast Threshold and Astronomical Visibility}
\author[Andrew Crumey]{Andrew Crumey\\Department of Humanities, Northumbria University, Newcastle upon Tyne NE1 8ST}

\maketitle

\begin{abstract}
The standard visibility model in light pollution studies is the formula of \citet{hecht}, as used e.g. by \citet{schaefer}. However it is applicable only to point sources and is shown to be of limited accuracy. A new visibility model is presented for uniform achromatic targets of any size against background luminances ranging from zero to full daylight, produced by a systematic procedure applicable to any appropriate data set (e.g \citet{blackwell}), and based on a simple but previously unrecognized empirical relation between contrast threshold and adaptation luminance. The scotopic luminance correction for variable spectral radiance (colour index) is calculated. For point sources the model is more accurate than Hecht's formula and is verified using telescopic data collected at Mount Wilson by \citet{bowen}, enabling the sky brightness at that time to be determined. The result is darker than the calculation by \citet{garstang2004}, implying that light pollution grew more rapidly in subsequent decades than has been supposed. The model is applied to the nebular observations of William Herschel, enabling his visual performance to be quantified. Proposals are made regarding sky quality indicators for public use.
\end{abstract}
\begin{keywords}
light pollution -- telescopes -- history and philosophy of astronomy -- sociology of astronomy
\end{keywords}

\section{Introduction}
\subsection{Contrast threshold}
Determining the faintest star or extended object visible to the naked eye or with a telescope is a problem of interest in light pollution studies, the history of astronomy, and vision science. It is an issue of public concern and economic importance given the growth of recreational `dark sky parks' for amateur astronomy and the imposition of lighting ordinances to preserve the aesthetic quality of the night sky \citep{idsa}. This article will present a model applicable to uniform achromatic targets of any size, seen against background luminance levels ranging from total darkness to daylight, hence relevant to visibility problems in many areas. For low light levels it will be applied to historical astronomical data and shown to be more accurate than previous models.

Visibility is dependent on the luminance (equivalent to surface brightness) of the target object, $B_{\mathrm{t}}$, in comparison with that of the surrounding field, $B$. For an opaque object in front of a background, the contrast is defined as 
\begin{equation}
 C=\frac{B_{\mathrm{t}}-B}{B} \equiv  \frac{\Delta B}{B}.
\label{c_def}
\end{equation}
For a target viewed through a transparent screen (or an astronomical object viewed through the atmosphere) the portion covering the target contributes luminance $B$, hence $\Delta B = B_{\mathrm{t}}$. When the increment $\Delta B$ is at the threshold of visibility according to specified criteria, then $C$ is the threshold contrast. For a target of angular area $A$ one can also consider the illuminance (equivalent to apparent magnitude). If $\Delta B$ is in candelas per square metre (cd m\textsuperscript{-2}) and $A$ is in steradians then the apparent increment illuminance at the eye in lux (lx) is
\begin{equation}
\Delta I = A\Delta B.
\label{delta_i_def}
\end{equation}
The modelling problem is to find analytic expressions for threshold $\Delta B$ or $\Delta I$ as functions of $A$ and $B$. This will then indicate the limiting surface brightness or magnitude for objects seen with the naked eye against a sky with luminance $B$. Telescopic results can be found by applying standard optical formulae which take into account the changes in $A$ and $B$ imposed by magnification, and factors such as light loss. \citet{schaefer} investigated stellar limiting magnitude by adopting a pre-existing model formula for $\Delta I$ in terms of $B$ (his eq. 2) and applying modifications to those quantities. That work has been particularly important, so similarities and differences with the present treatment will be highlighted throughout this Introduction, which also outlines relevant aspects of psychophysics, vision science and photometry. The model will be constructed in Section 2 and shown to give an accurate representation of the laboratory data. In Section 3 it is applied to astronomical visibility and tested against historical limiting magnitude data collected at Mount Wilson, as well as the nebular discoveries of William Herschel. It is shown how the model can be used to make comparative predictions about the effect of sky brightness on telescope performance. The concluding section discusses the practical definition of `dark sky' for the purposes of visual astronomy.

\subsection{Blackwell's data}

The largest and most authoritative study of contrast threshold was that of \citet{blackwell}, whose data continue to be used in areas such as lighting engineering and road safety \citep{narisada}. They were used in a popular book for amateur astronomers by \citet{clark} to make graphical predictions of astronomical visibility. The approach to be taken here is instead analytic, giving rise to formulae relevant to a wide range of visibility problems, and applicable to data other than Blackwell's. Blackwell measured luminance in footlamberts, but the data will be presented here in modern units (1 fL = 3.426 cd m\textsuperscript{-2}). Blackwell investigated both positive and negative contrasts, but only the positive data (target brighter than surround) will be considered in this article.

In order that the data can be correctly interpreted it is necessary to consider the experimental method in some detail. A total of 19 highly trained female observers aged 19-26 with approximately 20/20 vision were employed specially for the project (serving also as data analysts). No gender effect has been reported in the literature, but the age, experience and motivation of the observers is significant. Observers viewed targets using unconstrained (direct or averted) binocular vision for effectively infinite time (i.e. such that doubling of exposure did not alter successful detection rates). Targets, viewed at effectively infinite range (over 15 m) were uniform achromatic (broadband white) discs of seven angular sizes ranging from 0.595 arcmin to six degrees. These were either projected or, for the smallest target, transilluminated. They were viewed against backgrounds ranging from 3426 cd m\textsuperscript{-2} down to zero, and at five levels of contrast (in relative proportion 0.24, 0.37, 0.55, 0.75, 1). Observers were always allowed to become fully adapted to the background luminance, and rest periods were given so as to avoid fatigue. A single session would consist of 320 presentations, and observers were not considered trained until they had participated in approximately 20 sessions, though the published data were based on a far higher level of experience (35-75,000 observations by each observer) resulting in `unusual sensitivity and gratifying stability of response'.

For every target and contrast level, each observer's probability of detection was found. For each observer's probability curve (approximately a normal ogive), a graphical procedure was used to extract the contrast value $M$ that would correspond to 50 per cent detection, chosen since it could be calculated with highest precision, and $\sigma$, the standard deviation of the normal probability integral. Results from 1500 probability curves were averaged, giving contrast threshold as a discrete function of target size and background luminance: table 7 of \citet{blackwell} summarises 90,000 observations by seven observers. Smooth curves were drawn as a best fit to these data points, and interpolations made, to produce the final values (table 8 of \citet{blackwell}). 

At ordinary light levels (for discs larger than 0.595 arcmin) a `forced-choice' procedure was used, in which the target would appear in one of eight possible positions (or not at all), and observers had to indicate where they thought the target was displayed using a selector switch at the end of the viewing period. At low light levels, requiring much longer viewing times, a two-valued forced-choice procedure was used (`yes-no'), in which the target was presented (or not) at the centre of the screen, and observers had to indicate whether they had seen it, again with a switch. Null targets enabled the effect of random guessing to be eliminated.

The 50 per cent detection level was merely a statistical normalisation: the threshold at any other detection probability $p$ could be found by applying a multiplier $f = 1 + (\sigma /M)z$, where $z$ is the normal distribution standard score for cumulative probability $p$ \citep{BlackwellBrightnessDiscriminationData}. It is interpreted as meaning that an observer, exposed many times to a threshold target under the conditions of the experiment, would be expected to respond correctly on 50 per cent of occasions. It does not mean that during a single observation the target should be visible for 50 per cent of the time, as suggested by \citet{clark1994}: an observer able to see the target for any period of time during every exposure would be expected to achieve a success rate of 100 per cent (assuming no mistaken responses or false positives). Nor does it mean that the observer would be 50 per cent confident of having seen the target, as supposed by \citet{schaefer}. Blackwell reported that, in general, observers were confident of having seen the target only in cases where the resulting detection probability was 90 per cent or greater, corresponding to $f = 1.62$, suggesting that thresholds should be multiplied by at least this much to give realistic values. \citet{blackwellblackwell} noted that subjects in forced-choice experiments could show a detection rate slightly better than chance, even when not consciously aware of having seen the target. Higher (poorer) thresholds are found if observers instead adjust the brightness until the target becomes just visible. In order to raise forced-choice thresholds to what they termed `common-sense seeing', Blackwell \& Blackwell proposed a multiplier 2.4. The application of an overall multiplier to laboratory data will here be termed `laboratory scaling', and can be used as a way of comparing data from studies performed under different experimental conditions.

Blackwell's data were extended to larger target sizes in \citet{taylorlarge1} and \citet{taylorlarge2}, and to all observer ages by \citet{blackwellblackwell}. Threshold was found to rise with increasing age: slowly up to about 45, then quite rapidly. This is due mainly to loss of transparency in the ocular media (Adrian 1989) rather than diminution of pupil size, as assumed by \citet{schaefer}. However it was found that to a very good approximation the shape of the threshold curve (on log axes) remained invariant, i.e. the effect of age is to introduce a further overall multiplier, in addition to laboratory scaling. Empirically determined age multipliers in \citet{blackwellblackwell} range from 1 for 50 per cent of 20-year-olds, to 6.92 to include 95 per cent of 65-year-olds.

The age multiplier is an example of a `field factor' \citep{taylorvis} constituting a departure from the laboratory conditions. More generally these may be associated with the target (e.g. non-circular shape, non-uniformity), viewing conditions (non-uniform background, glare sources) and observer (motivation, fatigue). The field factors contribute further multipliers to the contrast threshold. There may also be physical effects which objectively alter the stimulus, such as magnification or light loss in a telescope, needing to be taken into account as $A \rightarrow A^\prime$, $B \rightarrow B^\prime$. Hence in practice the threshold function is $FC(A^\prime,B^\prime)$ where $F$ - the product of all field factors and any laboratory scaling - has the effect of shifting the whole curve up or down on log axes. Thus the threshold function should in general be considered relative rather than absolute \citep{blackwellpsychophys}, but invariant in shape, implying a correlation between threshold at large and small target sizes (i.e., in astronomy, an equivalence between limiting stellar magnitude and limiting surface brightness).

This suggests two approaches to visibility modelling. One (here called `enumeration') is to attempt to quantify all the relevant field factors from physical data (or estimate them) and hence determine $F$ for a given observing situation. The other (`elimination') is to leave $F$ as an unknown variable, unless data are available that allow it to be determined, or calculations can be performed where it cancels out. \citet{schaefer} studied stellar visibility using enumeration, whereas the present article will treat targets of arbitrary size using elimination. Schaefer assumed that the personal factor of the observer (which he denoted $F_\mathrm{s}$) was  approximately 1, but that the detection probability could vary, whereas actually the latter is a fixed normalisation constant and the personal factor may vary considerably between individuals. The error was mathematically insignificant since the threshold is multiplied by the product of these factors.

\subsection{Photometric considerations}
\label{photometry}

Luminance can be defined as

\begin{equation}
B_{v} = K_{v}\int_0^\infty E(\lambda)v(\lambda)\mathrm{d}\lambda.
\label{luminance}
\end{equation}
where $E$ is the spectral radiance (e.g. Planck's equation if the source is a black body) and $v$ is a sensitivity function (of finite support) with associated normalisation constant $K_v$. Examples of $v$ are the International Commission on Illumination (CIE) 1924 photopic luminous efficiency function for a 2-degree field (here denoted $V_\mathrm{ph}$, with $K_\mathrm{ph} = 683$ lm/W) and the CIE 1951 scotopic function, $V_\mathrm{sc}$ (with $K_\mathrm{sc} = 1700$ lm/W), both tabulated in \citet{stockman}. One could also choose the normalised passbands of the Johnson-Morgan system ($S_\mathrm{U}$, $S_\mathrm{B}$, $S_\mathrm{V}$), tabulated in \citet{bessell}. Apparent $v$-magnitude can be defined differentially for targets of equal area as
\begin{equation}
m_{v}^1 - m_{v}^2 = 2.5\mathrm{log}(B_{v}^2/B_{v}^1).
\label{magnitude}
\end{equation}
Luminance in cd m\textsuperscript{-2} is conventionally defined using $V_\mathrm{ph}$ in Eq. \ref{luminance}, whereas magnitude in $\mathrm{mag}_\mathrm{V}$ is defined using $S_\mathrm{V}$ in Eq. \ref{magnitude}. Also of importance in light-pollution studies is the sensitivity function of the Unihedron Sky Quality Meter, $S_\mathrm{SQ}$. The similarity of these three sensitivity functions makes them effectively equivalent for most practical purposes (\citet{schaefer96}, \citet{cinzano}).

If the $v$-magnitude scale has zero-point $Z_v$ lx then $J$ lx is equivalent to apparent magnitude $m_v=2.5\mathrm{log}(Z_v/J)$, and $B$ cd m\textsuperscript{-2} is equivalent to surface brightness $\mu_v= 2.5\mathrm{log}(60^{4}(180/\pi)^{2}Z_v/B) \mathrm{mag}_v$arcsec\textsuperscript{-2}. Taking $Z_\mathrm{V} = 2.54\times10^{-6}$ lx \citep{allen} gives conversion formulae $m_\mathrm{V} = -2.5\mathrm{log}J - 13.99$, $\mu_\mathrm{V} = -2.5\mathrm{log}B + 12.58$. Henceforth magnitude can be assumed to be $V$-band unless indicated otherwise.

The darkest skies on Earth have a zenith luminance of approximately 22 mag arcsec\textsuperscript{-2} ($1.71 \times 10^{-4}$ cd m\textsuperscript{-2}), with the visible sky background on a clear moonless night being a combination (in descending order) of natural airglow, zodiacal light and unresolved starlight \citep{leinert}. Airglow typically accounts for about 60 per cent of zenith sky luminance \citep{leinert95}, varying with solar activity, and dominated by the 557.7 nm OI emission line which alone typically contributes around 20 per cent of total $V$-band sky brightness \citep{patat2008}. At urban sites the sky spectrum is dominated by tropospheric scattering of anthropogenic light; measurements by \citet{puschnig} near Vienna showed very large peaks at 546 nm and 611 nm (attributable to fluorescent street lamps) with smaller intervening peaks due to high-pressure sodium lamps. Zenith sky brightness was in the range 15-19.25 $\mathrm{mag_{SQ} arcsec^{-2}}$ (approximately $1.1 \times 10^{-1}$ to $2.2 \times 10^{-3}$ cd m\textsuperscript{-2}). As a general approximation one can take $2 \times 10^{-4}$ cd m\textsuperscript{-2} (21.83 mag arcsec\textsuperscript{-2}) as representative of a truly dark sky, though at a pristine site there may be regions of the sky that are darker than this \citep{duriscoe}.

Human vision at normal light levels is photopic, utilising the trichromatic cone cells whose density is greatest in the foveal region of the retina.  In very low light levels vision is scotopic, utilising the monochromatic rod cells whose density is greatest outside the fovea. 
There was formerly thought to be an abrupt switch between the two types of vision, though in fact there is a continuous transition at intermediate (mesopic) light levels as cone response lessens and rods become active. Hence it is misleading to speak of `day' and `night' vision. The transition is dependent on the particular visual task and prevailing conditions, so cannot be specified exactly, but as a working definition one can take the range of mesopic vision as 0.005 to 5 cd m\textsuperscript{-2} \citep{cie2010}, i.e. scotopic vision operates in conditions darker than about 18.3 mag arcsec\textsuperscript{-2}. For astronomical purposes, \citet{puschnig} estimated the limit as 18.9 $\mathrm{mag_{SQ} arcsec^{-2}}$ (approximately 0.003 cd m\textsuperscript{-2}).

Vision has `channels' for luminance and chromaticity; contrast can be defined with respect to either, but in scotopic vision only the luminance channel operates. If an observer is able to detect colour then this indicates cone activity, so nocturnal astronomical observation often involves mesopic rather than scotopic vision. Variable star observers generally restrict magnitude estimates to stars at least two magnitudes brighter than threshold, using direct rather than averted vision, which gives more reliable estimates since it minimises rod contribution \citep{hallett}. \citet{schaefer96} found from a survey of experienced observers that in telescopic stellar viewing, scotopic vision operates at no more than about one magnitude above threshold. Mesopic photometry would be important for a proper treatment of supra-threshold magnitude estimates or visibility under severe light pollution (including twilight). However this article is concerned with threshold rather than brightness perception; and although the model will cover achromatic vision across the entire luminance range, the astronomical applications will be restricted to scotopic vision, i.e. targets within about one magnitude of threshold against a background no brighter than about $3 \times 10^{-3}$ cd m\textsuperscript{-2} (18.9 mag arcsec\textsuperscript{-2}), to which the observer is assumed fully adapted.

Since photometric units are usually defined by photopic sensitivity, one might query the validity of low-level contrast thresholds measured in this way. However, suppose the background and target
have the same relative spectral radiance, i.e. $E$, $\alpha E$ respectively (as in Blackwell's experiment), with the background having luminance $B$ given by Eq. \ref{luminance}. Then the target has luminance $B_{\mathrm{t}} = \alpha B$, and from Eq. \ref{c_def} the contrast is $C = \alpha - 1$, independent of the sensitivity function used to define the photometry \citep{walkey}. There is, however, a dependency on $E(\lambda)$ which must be taken into account. Define

\begin{equation}
\rho_{E} = \frac{K_\mathrm{sc}\int_0^\infty E(\lambda)V_\mathrm{sc}(\lambda)\mathrm{d}\lambda}{K_\mathrm{ph}\int_0^\infty E(\lambda)V_\mathrm{ph}(\lambda)\mathrm{d}\lambda}.
\label{rho}
\end{equation}
This is the `S/P ratio'  \citep{cie2010} which characterises the relative output of a light source with spectral radiance $E$ as measured with respect to scotopic or photopic luminosity. A source with photopic luminance $B$ has scotopic luminance $\rho B$. Thus two sources could be measured as having equal (photopic) luminance, but one with lower S/P ratio will have less output (appear dimmer) at scotopic levels. Spectral radiance can also be characterised by correlated colour temperature (CCT), defined as the temperature of a black-body radiator whose perceived colour most closely resembles that of the light source. There is no general relationship between S/P ratio and CCT since light-sources are not in general black-body, though incandescent lamps are a very close approximation, and stars somewhat less so (due to absorption lines).

Of practical importance is where target and background are of differing relative spectral radiance, which is generally the case for objects seen against the sky. Suppose the target and background have spectral radiances $E_{\mathrm{\tau}}(\lambda)$, $E_{0}(\lambda)$ (not necessarily black-body), S/P ratios $\rho_{\mathrm{\tau}}$, $\rho_{\mathrm{0}}$, and the target luminance is measured as $B_{\mathrm{\tau}} = K_\mathrm{ph}\int_0^\infty E_{\mathrm{\tau}}(\lambda)V_\mathrm{ph}(\lambda)\mathrm{d}\lambda$, against background $B_{0} = K_\mathrm{ph}\int_0^\infty E_{0}(\lambda)V_\mathrm{ph}(\lambda)\mathrm{d}\lambda$. Then the scotopic luminance would be the same as a Blackwell target of measured (photopic) luminance $B_{\mathrm{t}} = (\rho_{\tau}/\rho_{2850})B_{\tau}$ against a background $B = (\rho_{0}/\rho_{2850})B_{0}$. We also have $B_{\mathrm{t}} = \alpha B$, i.e. $\alpha = (\rho_\tau B_\tau)/(\rho_0 B_0)$. Suppose that the target is at threshold, so $C = \alpha - 1$. Then the contrast of the actual target against its background is

\begin{equation}
C_{\tau} = \frac{B_{\mathrm{\tau}}-B_0}{B_0} = \frac{(\rho_{2850}/\rho_\tau)B_{\mathrm{t}}}{(\rho_{2850}/\rho_0)B_0} - 1 = \frac{\rho_0}{\rho_\tau}(C+1) - 1.
\label{correction}
\end{equation}
Hence in general there must be a correction to threshold values, though no correction is needed as long as target and background have the same relative spectral radiance.

Blackwell used incandescent light sources reported as having colour temperature 2850K \citep{tousey}, i.e. equivalent to CIE 1931 Standard Illuminant A. One must also take account of the spectral reflectance of the white screen, but this can reasonably be assumed constant for all visible wavelengths, in which case the light can be supposed to have been black-body radiation to a very good approximation. Evaluating Eq. \ref{rho} using Planck's equation and the tabulated luminous efficiency functions gives $\rho_{2850} = 1.408$. \citet {kth} found thresholds using incandenscent lamps with colour temperature 2360 K \citep{tousey}. If both teams measured a (photopic) luminance at some equal value then the scotopic luminance of Blackwell's light would be greater by a factor $\rho_{2850}/\rho_{2360} = 1.220$. To compare luminance measurements between the two experiments one would need to make this correction at scotopic levels. A correction would also be required at mesopic levels, but photometry for that case is not uniquely defined (\citet{rea}, \citet{cie2010}). By calculating S/P ratios for black-body temperatures $2000 \leq T \leq 50,000$ K, one finds the approximation
\begin{equation}
\rho_T =  (5.738\times10^6)/T^2 - (8.152\times10^3)/T +3.564,
\label{rhoapprox}
\end{equation}
which is accurate to within about 1 per cent across the range.

 Let $m_\mathrm{V}$, $m_\mathrm{B}$ denote the $V$- and $B$-magnitudes of a source with spectral radiance $E$. The $(B-V)$ colour index, $m_\mathrm{B} - m_\mathrm{V}$, is a further way of characterising $E$ in addition to $\rho_E$, though there is no general relationship between the two quantities. Let $m_\mathrm{ph}$, $m_\mathrm{sc}$ denote magnitudes for the same source with respect to $V_\mathrm{ph}$ or $V_\mathrm{sc}$ (with arbitrary zero-points $Z_\mathrm{ph}$, $Z_\mathrm{sc}$), and let subscript $e$ denote either $ph$ or $sc$. From the definitions (Eqs. \ref{luminance}, \ref{magnitude}, \ref{rho})

\begin{equation}
m_\mathrm{sc}-m_\mathrm{ph} = -2.5\mathrm{log}\rho_E + C_\mathrm{sc-ph},
\label{sc-ph}
\end{equation}

\begin{equation}
m_\mathrm{V}-m_\mathrm{e} = -2.5\mathrm{log}\left(\frac{\int_0^\infty E(\lambda)S_\mathrm{V}(\lambda)\mathrm{d}\lambda}{\int_0^\infty E(\lambda)V_{e}(\lambda)\mathrm{d}\lambda}\right) + C_\mathrm{V-e} - L_\mathrm{V-e},
\label{mmph}
\end{equation}
where $C_{x-y} = 2.5\mathrm{log}(Z_x/Z_y), L_{x-y} = 2.5\mathrm{log}(K_x/K_y)$. A natural zero-point choice is

\begin{equation}
C_\mathrm{sc-ph} = L_\mathrm{sc-ph} = 2.5\mathrm{log}(1700/683) = 0.990.
\label{cscph}
\end{equation}
One could calculate $m_\mathrm{V}-m_{e}$ using test functions for $E$ (e.g. black-body), or else by using stellar spectra, in which case there could also be a correction for atmospheric absorption and reddening. One could express the result in terms of colour index, in the first case using the same test functions to find an approximate relation, or in the second using the colour indices for the selected stellar spectra. \citet{schaefer96} calculated $m_\mathrm{V}-m_\mathrm{ph} = 0.042(m_\mathrm{B} - m_\mathrm{V})$ using the first method, while \citet{steffey} used stellar spectra for classes B0-M2 and presented graphical data showing a good straight-line fit $m_\mathrm{V}-m_\mathrm{sc} = 0.04-0.29(m_\mathrm{B} - m_\mathrm{V})$. Together these suggest

\begin{equation}
m_\mathrm{sc}-m_\mathrm{ph} \approx 0.25(m_\mathrm{B} - m_\mathrm{V}), 
\label{approx0.25}
\end{equation}
with due caution because of the differing methods used. There have also been empirical studies of supra-threshold (i.e. mesopic) visual magnitude estimates, $m_\mathrm{vis}$, seeking a linear relation $m_\mathrm{vis} - m_\mathrm{V} = a + b(m_{B} - m_\mathrm{V})$. By continuity and Eq. \ref{approx0.25} one expects $b \la 0.25$, lowering as rod contribution lessens. The surveys by \citet{landis}, \citet{bailey} and \citet{collins} support $a \approx 0$, $b \approx 0.22$, while \citet{stanton} used 650 specially acquired observations from 63 observers to obtain $a = 0$, $b=0.21$.

\citet{flower} found effective temperature versus colour index based on a large sample of stellar spectra. Relating his table 3 to Eq. \ref{rhoapprox} (and writing $m_{B} - m_{V}=c$) yields
\begin{eqnarray}
\mathrm{log}\rho_c = -0.05905c^6 + 0.1674c^5 - 0.06563c^4 \nonumber\\ - 0.1843c^3 + 0.2031c^2 - 0.1802c + 0.4447.
\end{eqnarray}
A linear approximation is
\begin{equation}
\mathrm{log}\rho_c = -0.1094c + 0.4378,
\label{bvlinear}
\end{equation}
which is accurate to within 5 per cent for $-0.17 \leq c \leq 1.65$. With Eqs. \ref{sc-ph} and \ref{cscph} this gives

\begin{equation}
m_\mathrm{sc}-m_\mathrm{ph}= 0.27(m_\mathrm{B} - m_\mathrm{V}) - 0.10,
\end{equation}
consistent with Eq. \ref{approx0.25}.

The colour correction for stars relative to laboratory sources can now be calculated. Let $T$ be the colour temperature of a (black-body) laboratory point source and $T_*$ be the effective temperature of a star (assumed to be black-body to good approximation), and let $B$, $B_*$ be their respective (photopic) luminances. Suppose they have the same scotopic luminance, i.e. $\rho_T B = \rho_{T_*} B_*$, and denote their $V$-magnitudes $m_T$, $m_*$. Let $c$ be the colour index of the star, i.e. $\rho_{T_*} \equiv \rho_c$. Then (from Eq. \ref{magnitude})

\begin{equation}
m_* - m_T = -2.5\mathrm{log}\left(\frac{\rho_T}{\rho_c}\right).
\label{m*-m}
\end{equation}
Using Eq. \ref{bvlinear} and the previously calculated values of $\rho_T$, the corrections for the laboratory temperatures of \citet{blackwell} and \citet{kth} can then be written

\begin{equation}
m_* - m_{2850} = 0.72 - 0.27(m_\mathrm{B} - m_\mathrm{V}),
\label{blackwellcorr}
\end{equation}

\begin{equation}
m_* - m_{2360} = 0.94 - 0.27(m_\mathrm{B} - m_\mathrm{V}).
\label{kthcorr}
\end{equation}
Note that if $m_1$, $m_2$ are the thresholds for colour indices $c_1$, $c_2$ then
\begin{equation}
m_1 - m_2 = 0.27(c_2 - c_1).
\label{starcorr}
\end{equation}

\citet{schaefer} calculated Eq. \ref{kthcorr}; his $F_\mathrm{c} = F_\mathrm{o}F_\mathrm{v}$ is equivalent to $\rho_{2360}/\rho_c$, though he worked in terms of photon count, in which case photonic passbands should be used \citep{bessell2012}. He gave limited details of the calculation but stated the approximate result $1- (m_{B} - m_\mathrm{V})/2$, which is contradicted by the present analysis (a quotient 4 would be acceptable). He appears not to have used it except in the derivation of his eq. 18, where he assumed a uniform value $\rho_{2360}/\rho_c = 0.5$ for stars. Schaefer stated that the correction should be applied to background as well as target, but this is only valid if the background is (approximately) black-body at visible wavelengths, and the night sky is not, due to airglow (and any anthropogenic light pollution). The colour index of the night sky is similar to that of the sun, but that does not imply similarity of S/P ratio.

The spectral radiance of the sky is variable: at a given site there will, for example, be temporal changes in airglow, variation of tropospheric scattering with zenith distance, and variation of zodiacal light with ecliptic latitude. Zodiacal light is scattered sunlight, and its spectrum is almost identical to the sun's at visible wavelengths (Fig. 38 of \citet{leinert}); integrated starlight has approximately the same spectrum over that range (Fig. 1 of \citet{leinert}). Hence the moonless zenith sky without light pollution can reasonably be approximated by a 5500 K black-body spectrum together with airglow lines in some relative percentage of luminance. Taking the typical airglow spectral data given in table 13 of \citet{leinert}, one finds $\rho_\mathrm{sky}$ ranging from 0.79 (100 per cent airglow) to 2.26 (0 per cent), with a typical figure (60 per cent) being 1.38. This is very close to the S/P ratio of Blackwell's light sources (1.41, equivalent to 58 per cent airglow). It is therefore reasonable to take the spectral radiance of Blackwell's background as a sufficiently good approximation of a typical moonless night sky in the absence of light pollution.

\subsection{The data of Knoll et al}
\label{knolldata}

\begin{figure}
\includegraphics[width=84mm]{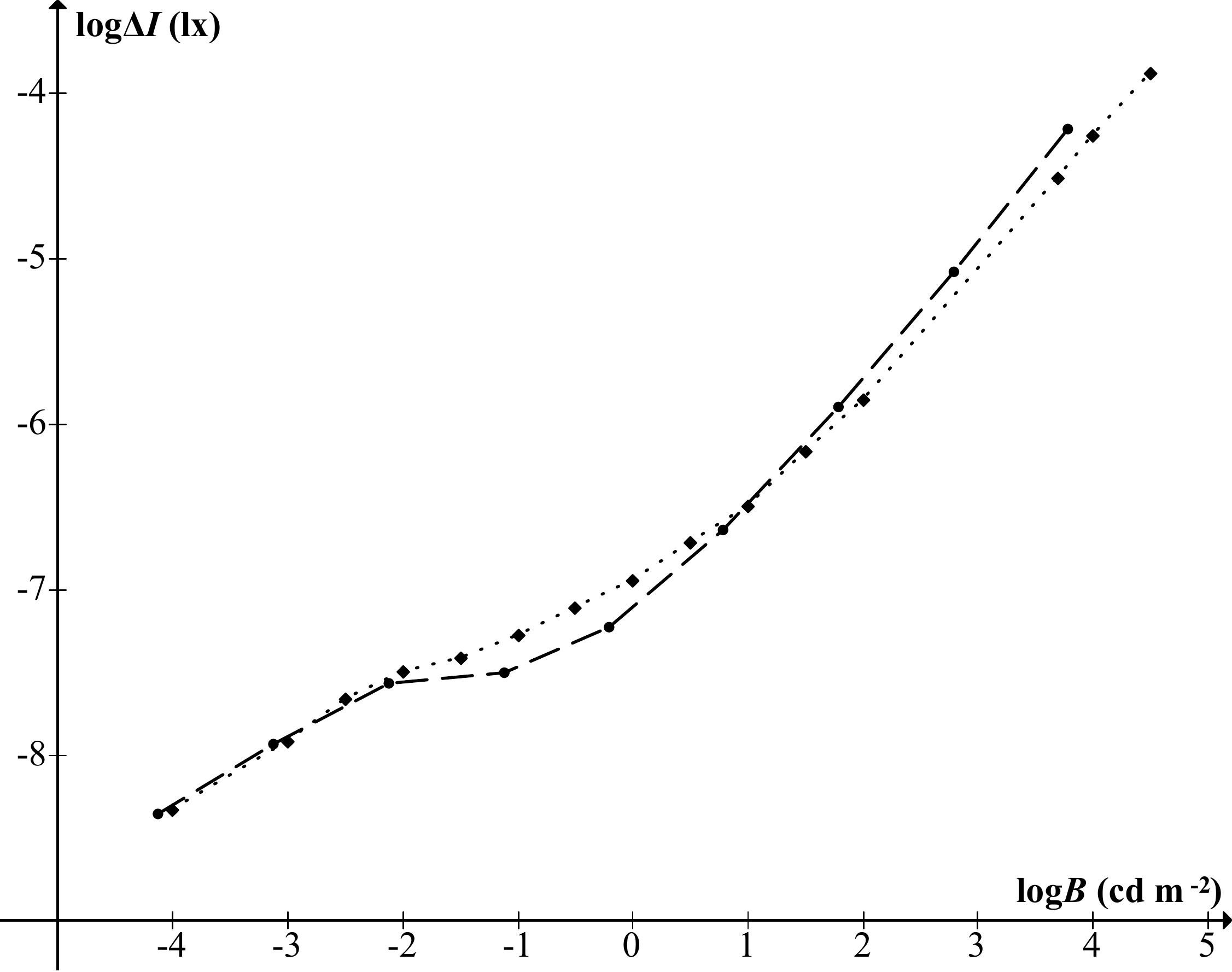}
\caption{Point-source threshold data from \citet{kth} (dotted line) and \citet{blackwell} (dashed), the latter having been adjusted as explained in Section \ref{knolldata}.}
\label{kthvblackwell}
\end{figure}

The point-source visibility study of \citet{kth} is of special interest because of its role in subsequent astronomical applications. Five young experienced observers used binocular vision to view a projected target of approximate diameter 1 arcmin. Each observer was given unlimited time to raise and lower the target brightness to find the level at which it was just visible. Thresholds were presented as equivalent increments ($\Delta I$) for an opaque target. To compare the results with the point-source thresholds of \citet{blackwell}, the latter's scotopic luminance values (for background and increment) must be multiplied by 1.220 because of the differing colour temperatures, and a further overall multiplier, $l$, must be applied to the increments because the method of adjustment yields higher thresholds than forced-choice with 50 per cent detection probability. \citet{tousey} proposed $l = 2$, because this was close to the Blackwell normalisation multiplier $f$ for forced-choice detection probability close to 100 per cent, though in the adjustment procedure the concept of detection probability is strictly meaningless. In fact one finds that $l=1.814$ brings the scotopic data ($-4 \leq \mathrm{log}B \leq -2$) into almost exact agreement (Fig. \ref{kthvblackwell}). Adjustment would also be required at mesopic levels, though some discrepancy would likely remain (as also at photopic levels), attributable to the differing experimental procedures, though not relevant to the astronomical situations considered in this article. For scotopic point-source thresholds (i.e. stars at night) the two data sets are effectively equivalent, though Blackwell's data are to be preferred as the more authoritative.

\subsection{Visibility models}

The earliest visibility models were for point-sources. \citet{langmuir} proposed $\Delta I \propto \sqrt{B}$, but this was poor. \citet{kth} offered
\begin{equation}
\Delta I = c(1+KB)^{1/2},
\label{kth}
\end{equation}
where $c = 1.076 \times 10^{-9}$ and $K = 10^{5}$ for $\Delta I$ in lx and $B$ in cd m\textsuperscript{-2}. Eq. \ref{kth}, like the Langmuir-Westendorp equation, took no account of the distinction between photopic and scotopic vision. In response, \citet{hecht} gave a formula derived from his own photochemical theory of retinal function \citep{hecht1934}, with two discontinuous branches:
\begin{equation}
\Delta I = c(1+ (KB)^{1/2})^{2},
\label{hecht}
\end{equation}
where (c,K) = $(1.706\times10^{-9}, 1.259\times10^{3})$ for $B\leq1.645\times10^{-2}$ cd m\textsuperscript{-2}, and (c,K) = $(4.808\times10^{-8}, 1.259\times10^{-1})$ for $B\geq1.645\times10^{-2} $ cd m\textsuperscript{-2}. This offered a better overall fit, especially in the photopic range. \citet{tousey} studied the visibility of stars in daylight and introduced a new empirical formula for $\Delta I$ which need not be considered further here, while \citet{weaver} studied night-time stellar visibility making use of Hecht's formula. That same formula was used by \citet{garstang1986} and then by \citet{schaefer}, whose work has formed the basis for most subsequent treatments, including the proposed extension of the model to finite target sizes by \citet{garstang1999}, and the light-pollution study of \citet{cinzano2001}. Eq. \ref{hecht} is also the basis for online limiting magnitude calculators widely used by amateur astronomers \citep{unihedron}.

\begin{figure}
\includegraphics[width=84mm]{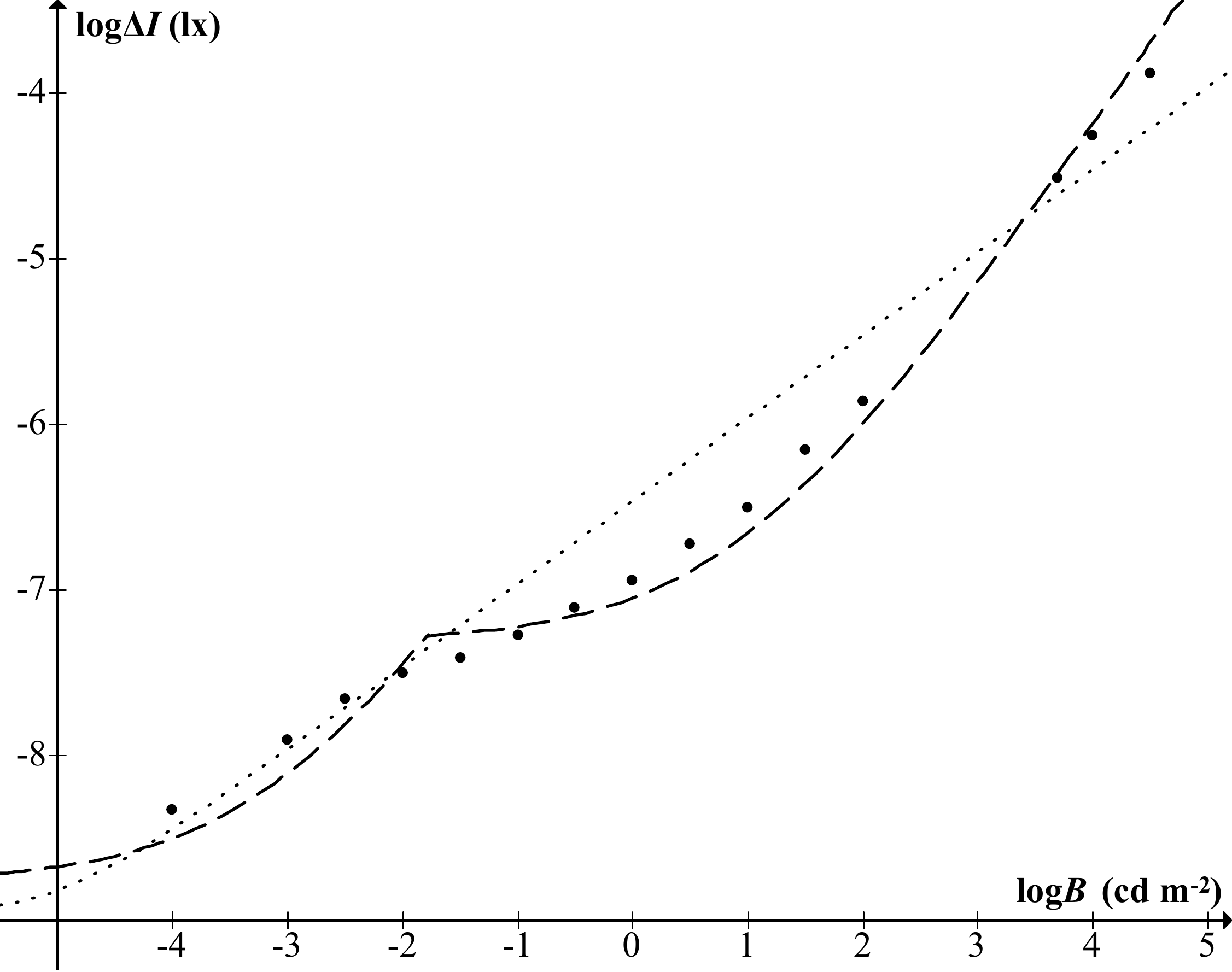}
\caption{Data from \citet{kth} together with their model (Eq. \ref{kth}, dotted) and that of \citet{hecht} (Eq. \ref{hecht}, dashed).}
\label{kthvhecht}
\end{figure}

While it has long been appreciated that Hecht's photochemical theory was invalid \citep{westheimer}, and that there is not really a discontinuous break between photopic and scotopic vision, what has not been noticed is that for the luminance range relevant to astronomical observation, Hecht's formula was actually inferior to the one by Knoll et al which it was supposed to replace. Fig. \ref{kthvhecht} shows the mean data from \citet{kth} converted to modern units, together with Eqs. \ref{kth} and \ref{hecht}. An equivalent graph was presented in the original units by \citet{hecht}. While it is evident that Hecht's model is greatly superior for photopic vision, it can be seen that this is not so in the scotopic (and lower mesopic) region. For $\mathrm{log}B$ in the range -1.5 to -4 (16.33 to 22.58 mag arcsec\textsuperscript{-2}) the data form a compressive curve whereas Hecht's curve is accelerating. The straight line of Knoll et al is better, but the new model presented here will be seen to be of the correct shape, providing a more accurate estimate of stellar visibility.

\subsection{Astronomical visibility factors}

\subsubsection{Viewing time}
Astronomical observations are often enhanced by long viewing times; \citet{clark1994} cited O'Meara's visual recovery of Halley's Comet after 1 to 2 hours, claiming it demonstrated a long-term integration property of the visual system. However saccades limit fixation time to no more than about a second, comparable with the maximum integration time of retinal cells. \citet{bishop} measured the shortest viewing times such that telescopic targets appeared undimmed compared with unlimited exposure, using the 0.61m telescope at Mauna Kea, and found times of 1.03 s or less. Confusion over Blackwell's definition of detection probability has led to an incorrect assumption that it is related to exposure time \citep{schaefer}, however laboratory experiments have shown that long viewing times generally degrade rather than enhance performance \citep{mackworth}. The benefit in astronomy can be explained by atmospheric variability; planetary observers are familiar with moments of best seeing, but what is less generally appreciated are fluctuations of transparency.

\subsubsection{Atmospheric turbulence}
\label{turbulence}
Air turbulence creates variations of refractive index manifested in seeing (image motion caused by tilting of wave-fronts) and scintillation (brightness variation caused by curved wave fronts focusing or defocusing starlight) \citep{dravins1}. The two effects are distinct, with major contributions from different atmospheric altitudes, and have little or no correlation, though aperture-dependency leads to an apparent correlation for naked-eye viewing, i.e. more noticeable scintillation on nights of poor seeing. A site with excellent seeing can nevertheless have high scintillation.
 
As will be explained in Section 2, at scotopic levels point sources are indistinguishable from extended targets up to about 10 arcmin in diameter. Hence seeing is important for high-magnification telescopic viewing, but has no effect on naked-eye viewing. Scintillation, on the other hand, has greatest effect for naked-eye viewing, and is of potential significance for threshold determinations with or without optical aid.

Scintillation occurs on multiple temporal scales, at all zenith angles, and can lead to sudden `flashes' with a brightening of 1 to 2 mag lasting a hundredth of a second, or lesser increase for longer \citep{ellison}. The visibility of brief flashes is dependent on their energy in relation to threshold \citep{blondel}; specific cases would require detailed calculation, but it can be seen that the general effect is that a very faint star may only be seen momentarily during many minutes of observation, and there may on occasion be a sighting of a star considerably fainter than the usual limit. Scintillation alters the colour of stars \citep{dravins2} and has a differential effect on point versus extended sources \citep{dravins1}. The effect of scintillation is averaged out by long integration times and large apertures, but human vision has a very short integration time and (for naked-eye viewing) a very small aperture, so that variations are potentially large. The phenomenon is caused by high altitude winds, so proximity to a main jet stream is expected to lead to higher scintillation: \citet{dravins3} noted the high rates measured at Mauna Kea and Paranal, and suggested that there would in general be greater scintillation along latitudes $\pm30$ degrees, and minima at the equator and poles.

The effect of seeing on telescopic views is dependent on aperture. The Fried parameter $r_0$ is the critical diameter above which resolving power is limited by the atmosphere rather than by the telescope's own diffraction \citep{fried}. In a sufficiently large telescope, a star produces a blurred disc of speckles (each of which is an Airy disc) with an approximately Gaussian profile. It is customary to quote the seeing $\theta$ as the disc's full width at half maximum (FWHM), though the actual image is larger. Since FWHM = $2\sigma\sqrt{2\mathrm{ln}2}$ for a Gaussian standard deviation $\sigma$, and since $\pm3\sigma$ will contain 97 per cent of the light of a Gaussian disc, one could take the actual width of the seeing disc as approximately $3/\sqrt{2ln2} = 2.55\theta$. To contain 100 per cent one could take $\pm3.3\sigma$, i.e. $2.80\theta$. \citet{schaefer} assumed the disc diameter to be equal to the quoted seeing (in his eq. 7), which may be true for small telescopes depending on how the seeing has been assessed by the observer. \citet{garstang2000} made the same assumption in his model, but applied it to large telescopes.

\subsubsection{Position, colour, shape, structure}

Zenith angle is a determinant of atmospheric extinction and sky brightness (eqs. 3 and 19 of \citet{schaefer}), as well as atmospheric reddening and scintillation. In the method of enumeration one requires absolute values for these, whereas for elimination it is sufficient to require that observations are all made under sufficiently equivalent conditions. Air mass affects point and extended sources slighty differently (see e.g. \citet{duriscoe} and references therein) and this would need to be taken into account if the most precise results were required, but will not be done here.

It has been shown that Blackwell's experiment at scotopic levels can be considered a good representation of 2850 K black-body sources against a sky with typical airglow and neligible light pollution. Targets with a spectral radiance very different from black-body (e.g. emission nebulae), or heavily light-polluted sky backgrounds, would require special treatment using the techniques of Section \ref{photometry}. If the concern is with finding limiting magnitude by the method of elimination, it is sufficient to assume that stars are all of approximately the same colour index (as was done by \citet{schaefer} and \citet{cinzano2001}), though not necessarily a specific value. For stars of specific colour index Eq. \ref{blackwellcorr} should be used.

Threshold for rectangular targets was investigated by \citet{lamar} who showed that area is a sufficient determinant for aspect ratios up to approximately 7. Hence the model should be adequate for elliptical targets with apparent eccentricity up to about this figure. Non-uniform targets will be treated approximately: it will be shown that realistic predictions can be made regarding the visibility of galaxies or the seeing discs of stars. The sky itself can be considered uniform in the immediate vicinity of a target, but field stars potentially introduce glare sources, such that a faint target may be invisible because of a brighter star in the vicinity. This glare effect can be treated in a standard way \citep{adrian} but specific problems of this type will not be considered here.

\subsubsection{Telescope use}
\label{telescopeuse}

Light loss in a telescope introduces a differential effect with respect to naked-eye viewing, and constitutes a stimulus modification. \citet{schaefer} assumed transmittance values according to telescope type, but it will be seen that if multiple observations are recorded under suitably controlled conditions (as was done by \citet{bowen}) then the transmittance (denoted $F_\mathrm{t}^{-1}$) can be deduced by elimination. Monocular vision through a telescope introduces another differential effect, though this is a modification of threshold rather than stimulus. \citet{lythgoe} measured contrast thresholds with left ($C_\mathrm{L}$), right ($C_\mathrm{R}$) and both eyes ($C$), finding the approximate relation $1.4C = 0.5(C_\mathrm{L} + C_\mathrm{R}$). Theoretical considerations suggested that the factor on the left should be $\sqrt{2}$, and if threshold is equal in both eyes then this means the monocular threshold is $\sqrt{2}$ times the binocular value. This is the factor that was assumed by \citet{schaefer} (his $F_\mathrm{b}$), though it was included incorrectly as a stimulus modification in his eq. 15 (i.e. as a multiplier of $B$), and this was repeated by \citet{garstang2000}.

Magnification produces an increase of target area and also in most cases reduction of retinal illumination, because the observer effectively views through an artificial pupil (i.e. the exit pupil of the instrument, which is usually smaller than the eye pupil). In some cases the Stiles-Crawford effect may need to be considered (i.e. the reduction in luminous efficiency of rays entering the eye obliquely). This is significant for photopic (and mesopic) vision, being attributable to directional sensitivity of cone cells, but \citet{flamant} found little or no directional sensitivity in rod cells, while \citet{vanloo} found a very small effect, but only for rays entering at the periphery of pupils larger than about 5mm. Hence they stated that the usual equation for the photopic effect could not be applied, though \citet{schaefer} proposed such an expression (his eq. 9) which incorrectly gives a non-zero value for all pupil sizes. In the present article the effect will be considered negligible.

Telescope use potentially introduces other differential factors relative to naked-eye viewing. Observers may be apt to use near rather than infinite eye focus when looking through an eyepiece, which may alter the effect of any ocular aberration. The telescope itself may suffer from aberration, and will show the viewer a much smaller apparent area of sky, set within a darker surround. There may be a difference of search procedure between naked-eye and telescopic viewing (e.g. finding known stars to assess naked-eye limit, then searching for hitherto unknown ones to assess telescopic limit). If the telescope is undriven then motion may be a factor. These will be assumed part of an overall telescopic field factor whose components can be split into magnification dependent ($F_\mathrm{M}$) or independent ($F_\mathrm{T}$) terms. If an approximate value is needed, it will be assumed that $F_\mathrm{T} = \sqrt2$, i.e. the only significant magnification-independent factor is the correction for monocular vision. $F_\mathrm{M}$ can be considered to be unity at low magnification in the absence of the Stiles-Crawford effect, but contributions could come from the use of interchangeable eyepieces of differing specification and quality, and at high magnification the point spread function of the eye will be significant \citep{watson}, with an exit pupil of 0.5mm usually being regarded as the limit below which diffraction in the eye begins to dominate \citep{jacobs}. This imposes a maximum useful magnification \citep{angers}, apart from the limitations imposed by seeing. In practice it should be sufficient to assume $F_\mathrm{M}=1$ up to some magnification beyond which there is no further improvement in threshold.

\subsubsection{Definition of threshold}

The activities of amateur astronomers can lie anywhere between science and recreational sport. If the latter, then the individual's concern with limiting magnitude may be to maximise it, whereas for science a main interest should be consistency of measurement. Scintillation in particular is a potential bonus from the recreational point of view, though a source of noise for science.

Threshold can be boosted in various ways: \citet{curtis} observed stars through a hole in a black screen (i.e. against a totally dark background) and in this way was able to see one of magnitude 8.3, and possibly one of magnitude 8.9, though his limit for stars seen against the sky was 6.5. Flickering is known to improve threshold \citep{kelly}: a rapidly operating shutter (e.g. a fan) adjusted to the optimum frequency of around 6 Hz, and placed in the line of sight (e.g. within a telescope), would produce some gain of magnitude. \citet{omeara} found that hyperventilation helped, consistent with the findings of \citet{connolly}, though excessive oxygen is damaging to the retina \citep{yamada}.

It has always been appreciated that the traditional naked-eye magnitude limit of 6 is merely approximate. \citet{weaver} commented on the magnitude limits of nineteenth-century naked-eye star catalogues, which ranged from 5.7 (Argelander) to 6.7 (Heis). The latter observer was renowned for his visual acuity, and his \textit{Atlas Coelestis} is unusual in including the galaxy M33 as a naked-eye object \citep{heis}. Gould's \textit{Uranometria Argentina} had a stated magnitude limit of 7 but modern photometry has shown the actual limit to be 6.5 \citep{gould}. As an example of exceptional eyesight, Weaver cited Meesters' ability to see stars to 6.9 mag. Weaver's study upheld a value of just over 6 for the typical dark-sky naked-eye limit, yet more recently there has been a substantial raising of achievement and expectation. The Bortle Scale \citep{bortle} suggests that for a Class 1 (`excellent') site the limiting magnitude should be `7.6 to 8.0 (with effort)' and for Class 2 (`typical truly dark site') 7.1 to 7.5. \citet{schaefer} reported O'Meara's extraordinary ability to see stars as faint as 8.4 mag against the sky. Apart from unusual acuity or special observing techniques, such high limits may in many cases be explained by scintillation, with momentary glimpses being taken as typical threshold. Subjective estimates may not always be reliable or accurate; the survey by \citet{schaefer} yielded many responses (about half of the total) in which naked-eye limit was given only to the nearest whole number.

It is a matter of policy judgment whether visibility recommendations for the general public should be based on typical or extreme performance. For modelling, one requires a definition that most closely resembles the conditions of the laboratory data (Blackwell's experiment) and is not unduly sensitive to local effects or false positives. Probably the best way to achieve this in practice is the method of star counting, using designated areas close to zenith. Stars should be continuously visible (with direct or averted vision) for some extended period (seconds) rather than be seen to flash momentarily. The observer should be fully dark-adapted, with screening from terrestrial glare if necessary, and with a naked-eye view of the sky that is at least as large as a typical apparent field of view in a telescope (e.g. 50 degrees). For telescopic views the use of magnitude sequences (as described by \citet{schaefer}) is convenient, but should be consistent with naked-eye procedures. It will be implicitly assumed in subsequent discussion that thresholds effectively conform to these or similar criteria.

\section{The visibility model}
\subsection{Modelling strategy}

\begin{figure}
\includegraphics[width=84mm]{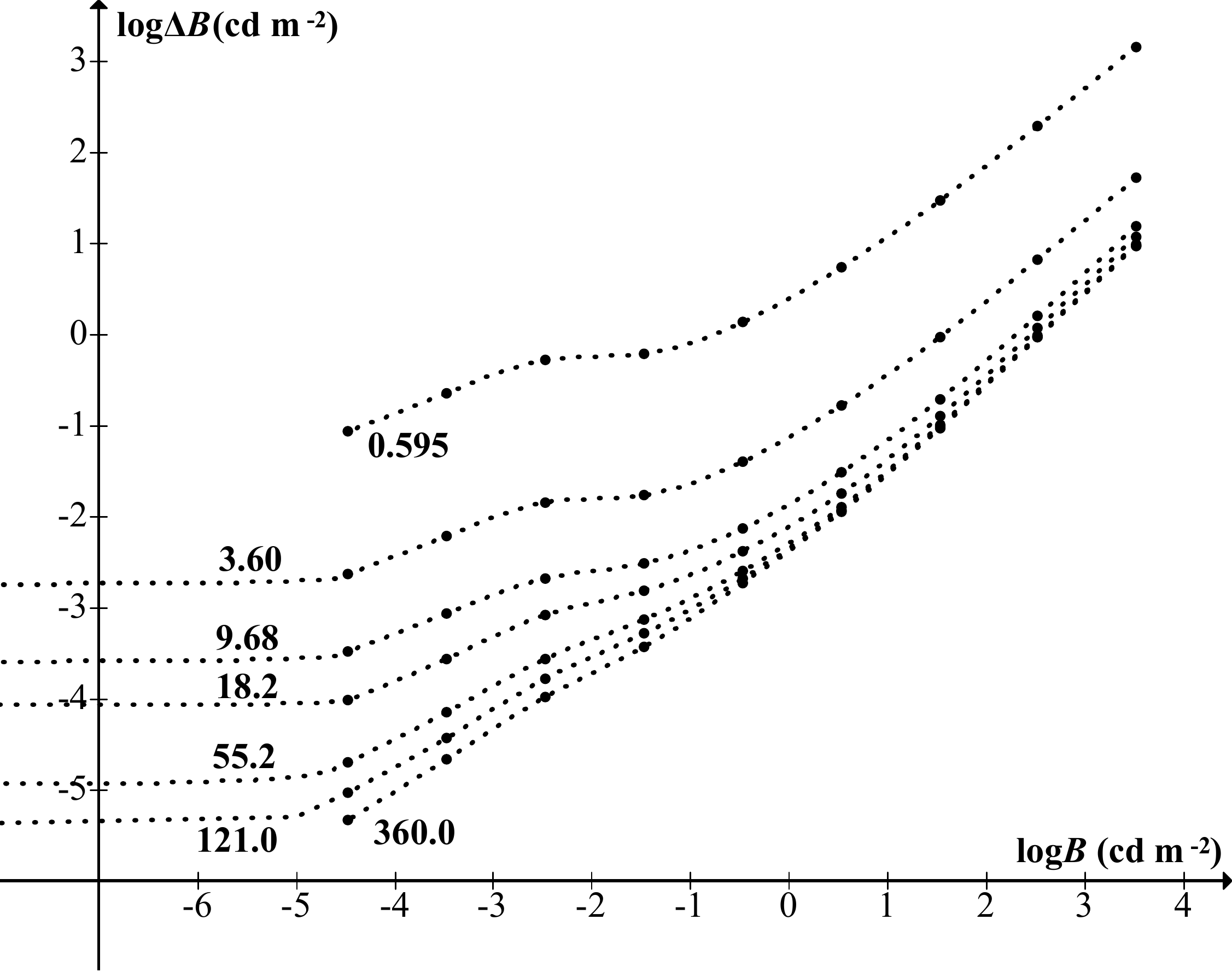}
\caption{Threshold increment versus background luminance for various target diameters (in arcmin). Data from tables 4 and 8 of \citet{blackwell}.}
\label{deltabvb}
\end{figure}

Fig. \ref{deltabvb} shows $\mathrm{log}\Delta B$ as a function of $\mathrm{log}B$ for various target sizes, using Blackwell's data. In daylight conditions the slope is approximately 1, i.e. $C$ = constant, which is Weber's Law. For extremely low $B$ the slope of the graph is zero, where $\Delta B$ has a non-zero limiting value attributable to neural noise (`dark light'). The curves indicate that a background $B \la 10^{-5}$ cd m\textsuperscript{-2} (25.08 mag arcsec\textsuperscript{-2}) is effectively zero for human vision, a finding also made by \citet{crawford}.

\begin{figure}
\includegraphics[width=84mm]{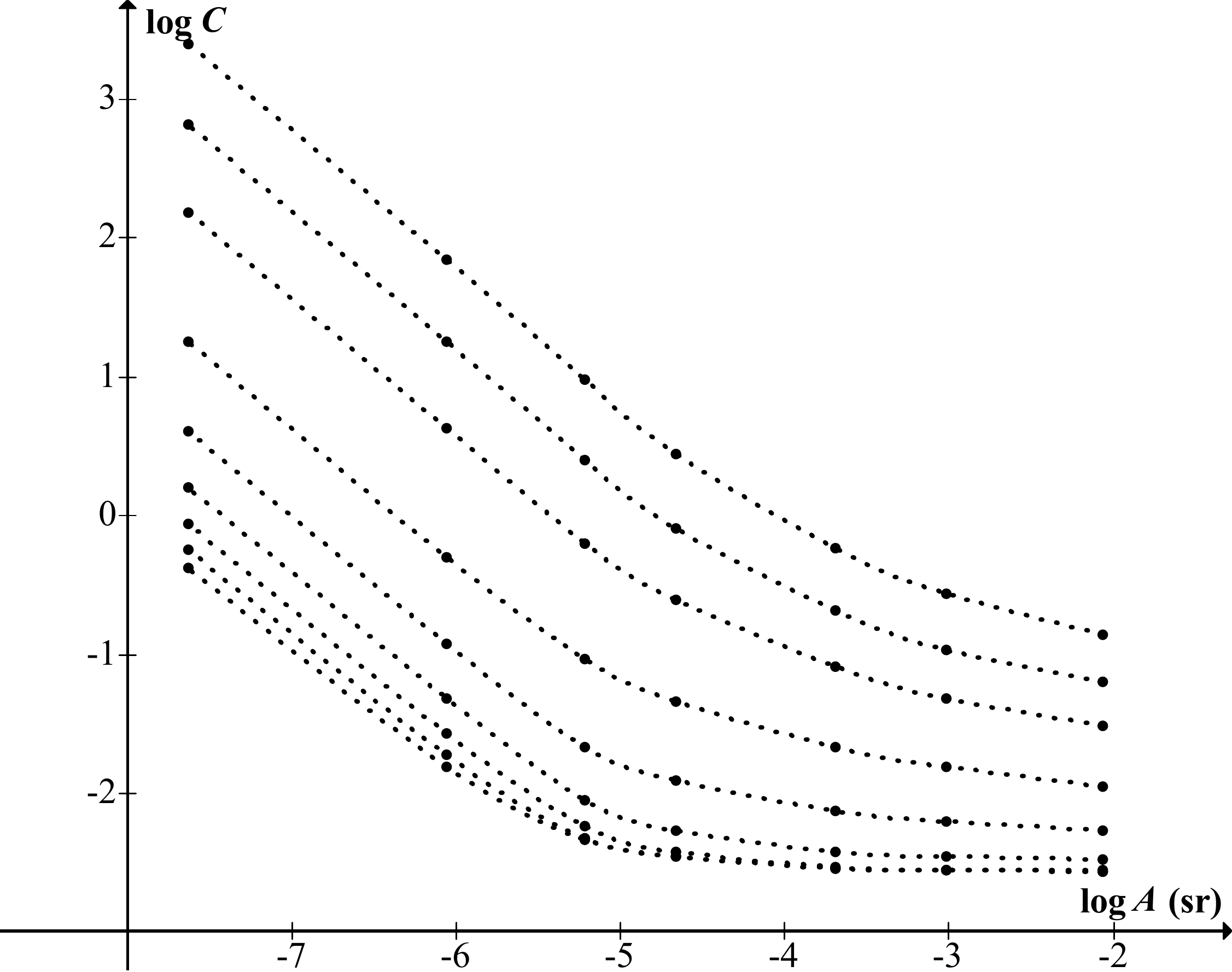}
\caption{Contrast threshold for different values of backround luminance $B$, from table 8 of \citet{blackwell}. Curves are at intervals of one log unit, from $3.426 \times 10^{-5}$ cd m\textsuperscript{-2} (top) to $3.426 \times 10^{3}$ cd m\textsuperscript{-2} (bottom).}
\label{cva}
\end{figure}

Fig. \ref{cva} shows $\mathrm{log}C$ as a function of $\mathrm{log}A$ for different levels of background luminance. For each $B$, the graph is asymptotic at both ends. For large $A$ the slope tends to zero, i.e. the contrast threshold approaches some limiting value $C_{\mathrm{\infty}}$ which is higher, and reached far more slowly, as $B$ decreases. For small $A$ the graph becomes a straight line of slope -1, i.e.
\begin{equation}
CA = R,
\label{ricco}
\end{equation}
where $R$ is a constant. This is Ricco's Law \citep{ricco}, and the maximum size for which it applies is sometimes called the Ricco area. The physiological intepretation is that the visual receptive field (corresponding to a number of receptor cells) sums the total energy received over its area, with a certain minimum energy being required in order to initiate a reaction. Both the Ricco area and the constant, $R$, become larger as the background luminance $B$ decreases. The significance in visual astronomy is that threshold targets subtending less than the Ricco area are indistinguishable from point sources, hence faint stars can be mistaken for nebulous objects and vice versa. This is reflected in the New General Catalogue \citep{dreyer}, where a number of entries are mis-identified stars. \citet{hubble} noticed an analogous effect occuring with threshold images on photographic plates, which he attributed to a combination of the photographic process and visual inspection.

Blackwell defined the `critical visual angle' graphically as the point where the threshold curve (reading left to right in Fig. \ref{cva}) begins to deviate from a slope of -1, so that Ricco's Law no longer holds. The more usual convention \citep{adrian} is to define the Ricco area $A_{\mathrm{R}}$ as the intersection of the asymptotes of the threshold curve, i.e.
\begin{equation}
A_{\mathrm{R}}= R/C_{\mathrm{\infty}}.
\label{riccoarea}
\end{equation}

The modelling strategy is first to find $R$ and $C_{\mathrm{\infty}}$ as functions of $B$, so that the asymptotes can be written as $C_{\mathrm{low}} = R(B)/A$ and $C_{\mathrm{high}}= C_{\mathrm{\infty}}(B)$. The full $C$ curve can then be obtained by smoothly piecing together the two asymptotes. \citet{koopman} suggested using an exponential joining function but did not obtain analytic expressions for the asymptotes. \citet{adrian} found asymptotic expressions by curve fitting and used the combined function $C = (C_{\mathrm{low}}^{2}+ C_{\mathrm{high}}^{2} )^{1/2}$, which was also adopted by the International Commission on Illumination \citep{cie}. A different model was offered by \citet{matchko}. All of these involved a large number of tuneable parameters.

The approach to be taken here is new, and is based on the suprising finding that $R$ and $C_{\mathrm{\infty}}$ are both simple functions of $B^{-1/4}$, across appropriate ranges of $B$. Model parameters are then specified by linear relations, in a systematic procedure that can be applied to any appropriate data set. The complete function will be $C = (C_{\mathrm{low}}^{q}+C_{\mathrm{high}}^{q})^{1/q}$, where $q$ is the only tuneable parameter in the model. 

\subsection{Point-source model}

\begin{figure}
\includegraphics[width=84mm]{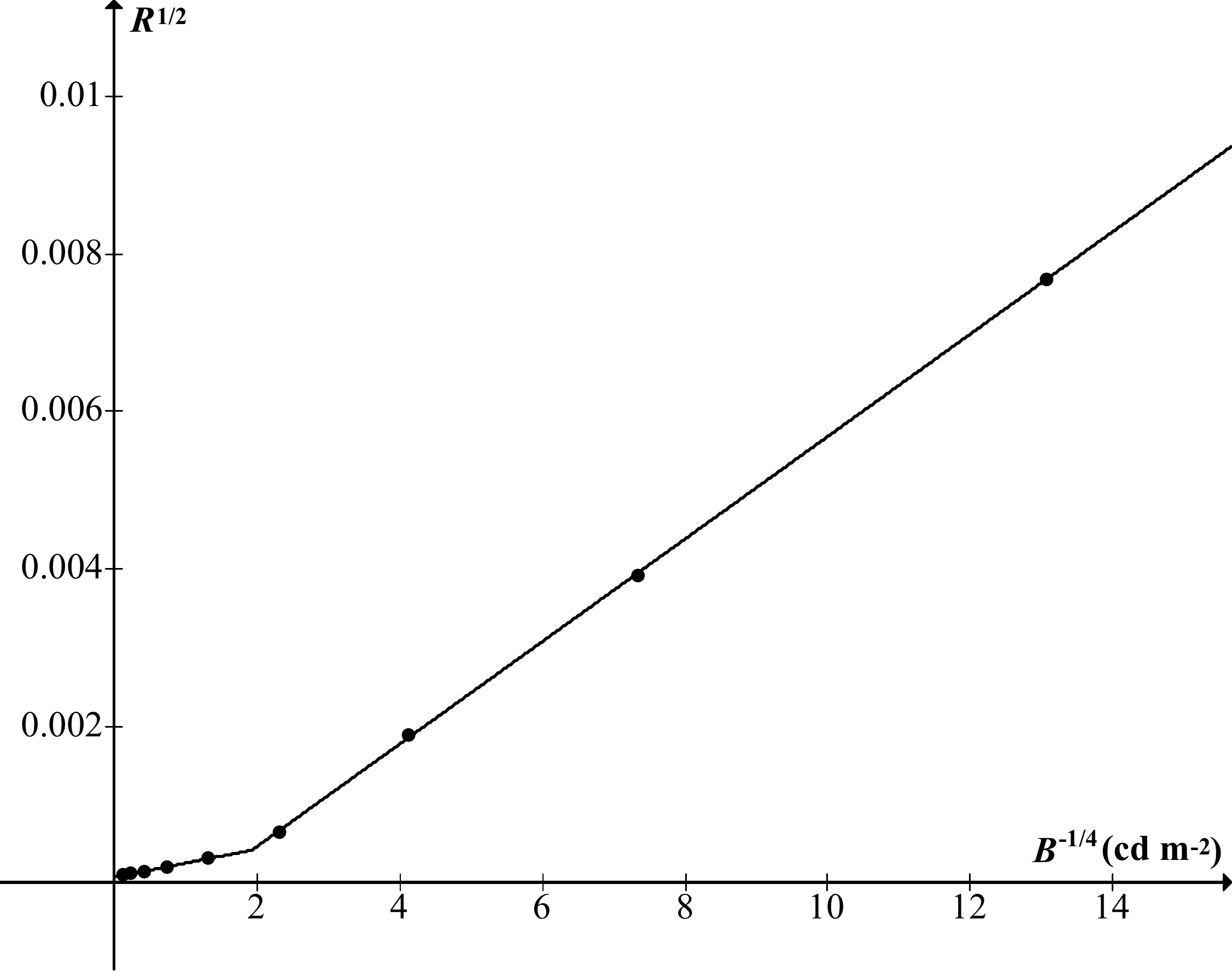}
\caption{$\sqrt{R} = \sqrt{CA}$ versus $B^{-1/4}$, using data from table 8 of \citet{blackwell} for target diameter 0.595 arcmin.}
\label{rbtable8}
\end{figure}

The asymptotic behaviour of $C$ for small area $A$ is obtainable from data for targets small enough to be effectively point sources, i.e. such that Ricco's Law (Eq. \ref{ricco}) is valid. One can calculate $R = CA$ from these data, then investigate the dependecy of $R$ on $B$. The form of Eqs. \ref{kth} and \ref{hecht} motivates the search for an empirical formula involving simple rational powers. A striking relation emerges when one plots $\sqrt{R}$ versus $B^{-1/4}$, as shown in Fig. \ref{rbtable8}. The graph consists of two linear sections roughly corresponding to photopic and scotopic vision, i.e.
\begin{equation}
R_{\mathrm{scot}}= (r_{1}B^{-1/4}+ r_{2})^{2},
\label{rscot}
\end{equation}
\begin{equation}
R_{\mathrm{phot}}= (r_{3}B^{-1/4}+ r_{4} )^{2},
\label{rphot}
\end{equation}
for constants $r_i$ obtainable by linear regression. The discontinuity between the two branches is a mathematical artefact rather than physiological fact: one assumes there must be a short but continuous bend joining the two straight sections. Hence the graph is really an almost-degenerate hyperbola $(\sqrt R - \sqrt R_{\mathrm{scot}})(\sqrt R - \sqrt R_{\mathrm{phot}}) = r_{5}\approx0.$ By the quadratic formula this is equivalent to
\begin{equation}
R = (\surd(a_{1}B^{-1/2}+ a_{2}B^{-1/4}+ a_{3}) + a_{4}B^{-1/4}+ a_{5})^{2},
\label{rhyp}
\end{equation}
for constants $a_i$. One can use the $r_i$ values from Eqs. \ref{rscot} and \ref{rphot} to obtain $a_i$ algebraically (on the assumption that $r_{5}= 0$), then use those $a_i$ values as an initial step in a Gauss-Newton algorithm to find best-fitting values for the hyperbola as a whole. (In fact to achieve convergence it is found necessary to omit the data point for $\mathrm{log}B=-1.465$). One then has two model versions: a simple two-branched form involving $r_i$, suitable for cases restricted to one or other visual regime, and a more complicated expression in $a_i$ that covers the entire range. The parameters are found to be
\begin{equation}
r_{1}=6.505\times10^{-4}, r_{2}=-8.461\times10^{-4},
\label{r1r2}
\end{equation}
\begin{equation}
r_{3}=1.772\times10^{-4}, r_{4}=7.167\times10^{-5},
\label{r3r4}
\end{equation}
with split-point $B=7.08\times10^{-2}$ cd m\textsuperscript{-2}, and
\[
a_{1} = 5.949\times10^{-8}, a_{2} =-2.389\times10^{-7}, a_{3} =2.459\times10^{-7},
\]
\begin{equation}
a_{4} =4.120\times10^{-4}, a_{5} =-4.225\times10^{-4}. 
\label{ai}
\end{equation}

\begin{figure}
\includegraphics[width=84mm]{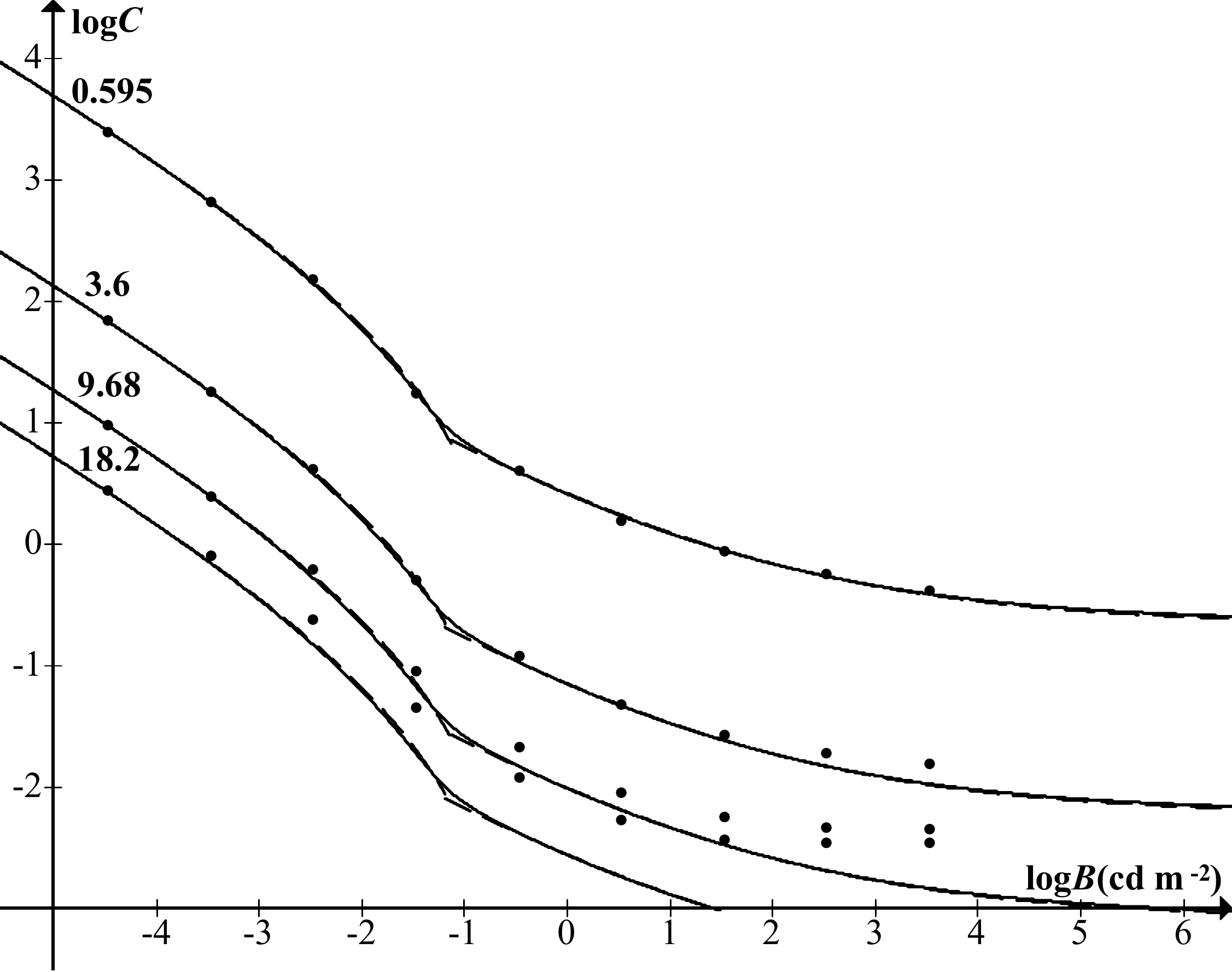}
\caption{Contrast threshold data from table 8 of \citet{blackwell} (target diameters labelled in arcmin), compared with $C = R/A$ calculated using Eqs. \ref{rscot}, \ref{rphot}, \ref{r1r2}, \ref{r3r4} (dashed lines) and Eqs. \ref{rhyp}, \ref{ai} (solid).}
\label{riccotableviii}
\end{figure}

In either case we can compare the resulting function $C = R/A$ with the original data set. This is seen in the uppermost curve of Fig. \ref{riccotableviii} which shows that both model versions fit the data very well, and apart from the transition region around $B = 7.08\times10^{-2}$ cd m\textsuperscript{-2} (15.5 mag arcsec\textsuperscript{-2}) they are virtually indistinguishable. One expects the point-source model to maintain accuracy across the range of validity of Ricco's Law, i.e. for target sizes up to the Ricco area. For daylight conditions this means a diameter of no more than about an arcminute, but in low light conditions the size increases. This is seen in the lower curves of Fig. \ref{riccotableviii}, which show that at low light levels the point-source model remains highly accurate for target diameters up to about 10 arcmin.

The relation shown in Fig. \ref{rbtable8} is also found in other point-source data sets, such as the mean data values given in table 7 of \citet{blackwell}, the data of \citet{kth} shown in Fig. \ref{kthvhecht}, or the data of \citet{siedentopf}. In fact the latter two data sets both show a smooth short bend between the asymptotes. One can apply the same procedure to any of these sets to obtain $r_i$ and $a_i$ values. The Blackwell table 8 values (Eqs. \ref{r1r2}, \ref{r3r4}, \ref{ai}) will be taken as definitive because of the authoritative nature of that data set, however it is also worth considering the model that arises from the data of Knoll et al, because of the significance of the Hecht formula, Eq. \ref{hecht}. One finds
\begin{equation}
r_{1}=7.310\times10^{-4}, r_{2}=-5.162\times10^{-4},
\label{kthr1r2}
\end{equation}
\begin{equation}
r_{3}=2.550\times10^{-4}, r_{4}=4.420\times10^{-5},
\label{kthr3r4}
\end{equation}
with split-point $B=5.21\times10^{-1}$ cd m\textsuperscript{-2} (13.3 mag arcsec\textsuperscript{-2}), and
\[
a_{1} = 6.112\times10^{-8}, a_{2} =-1.598\times10^{-7}, a_{3} =1.167\times10^{-7},
\]
\begin{equation}
a_{4} =4.988\times10^{-4}, a_{5} =-3.014\times10^{-4}. 
\label{kthai}
\end{equation}

The threshold increment illuminance at the eye is $\Delta I = A\Delta B = BR.$ Using Eqs. \ref{rscot} and \ref{rphot} we have
\begin{equation}
\Delta I_{\mathrm{scot}} = (r_{1}B^{1/4}+ r_{2}B^{1/2})^{2},
\label{deltaiscot}
\end{equation}
\begin{equation}
\Delta I_{\mathrm{phot}} = (r_{3}B^{1/4}+ r_{4}B^{1/2} )^{2},
\label{deltaiphot}
\end{equation}
while Eq. \ref{rhyp} gives the alternative form
\begin{equation}
\Delta I = (\surd(a_{1}B^{1/2}+ a_{2}B^{3/4}+ a_{3}B) + a_{4}B^{1/4}+ a_{5}B^{1/2})^{2}.
\label{deltaihyp}
\end{equation}

\begin{figure}
\includegraphics[width=84mm]{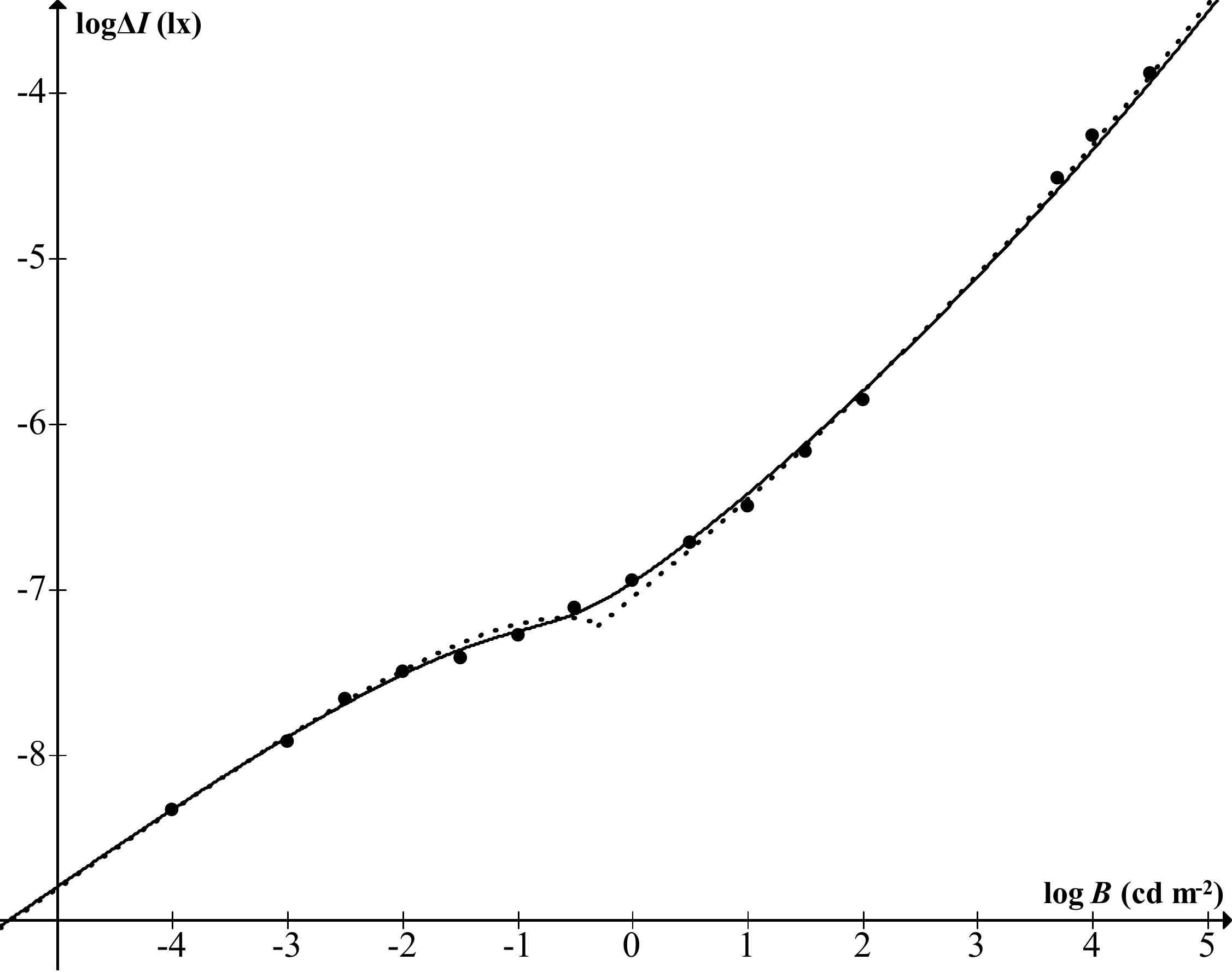}
\caption{Threshold increment illuminance $\Delta I$ versus background luminance $B$ for a point-source target: data from \citet{kth} together with two model versions. Dotted line is from Eqs. \ref{kthr1r2}, \ref{kthr3r4}, \ref{deltaiscot}, \ref{deltaiphot}; solid line is from Eqs. \ref{kthai}, \ref{deltaihyp}.}
\label{kthnewmod}
\end{figure}

Fig. \ref{kthnewmod} shows both versions together with the data of \citet{kth}, from which it can be seen that either form offers considerable accuracy. Comparison with Fig. \ref{kthvhecht} shows that the new model is substantially better than the ones previously proposed.

Henceforth only the Blackwell values (Eqs. \ref{r1r2}, \ref{r3r4}, \ref{ai}) will be used. It will be noted that the model (Eqs. \ref{deltaiscot}, \ref{deltaiphot}, \ref{deltaihyp}) does not have the correct asymptotic limit as $B \rightarrow 0$, since the threshold should tend to a non-zero value. This defect is not important for naked-eye astronomical observation, because of the natural brightness of the sky, but is relevant in telescopic observation, and will be addressed later.

\subsection{Full visibility model}

To construct the complete model it is necessary to find an analytic expression for $C_{\mathrm{\infty}}$, the threshold for large targets. At daytime luminance levels $C_{\mathrm{\infty}}$ is independent of $B$ (reflecting Weber's Law) and `large' is only a few arcminutes, but at low levels it rises above 6 degrees, the maximum target size in \citet{blackwell}. \citet{taylorlarge1} attempted to extend Blackwell's data in order to find $C_{\mathrm{\infty}}$ (at detection probability 0.5) for all background luminance levels, though he used a lower colour temperature (2360K), fixed viewing time (6 seconds) and a rather different methodology. His results were somewhat inconsistent with Blackwell's \citep{taylorlarge2}, with thresholds higher by a factor of approximately 2.2 for $B$ of the order of 1 cd m\textsuperscript{-2}, and approximately equal for $B$ of the order of $10^{-3}$ cd m\textsuperscript{-2} or less.

\begin{figure}
\includegraphics[width=84mm]{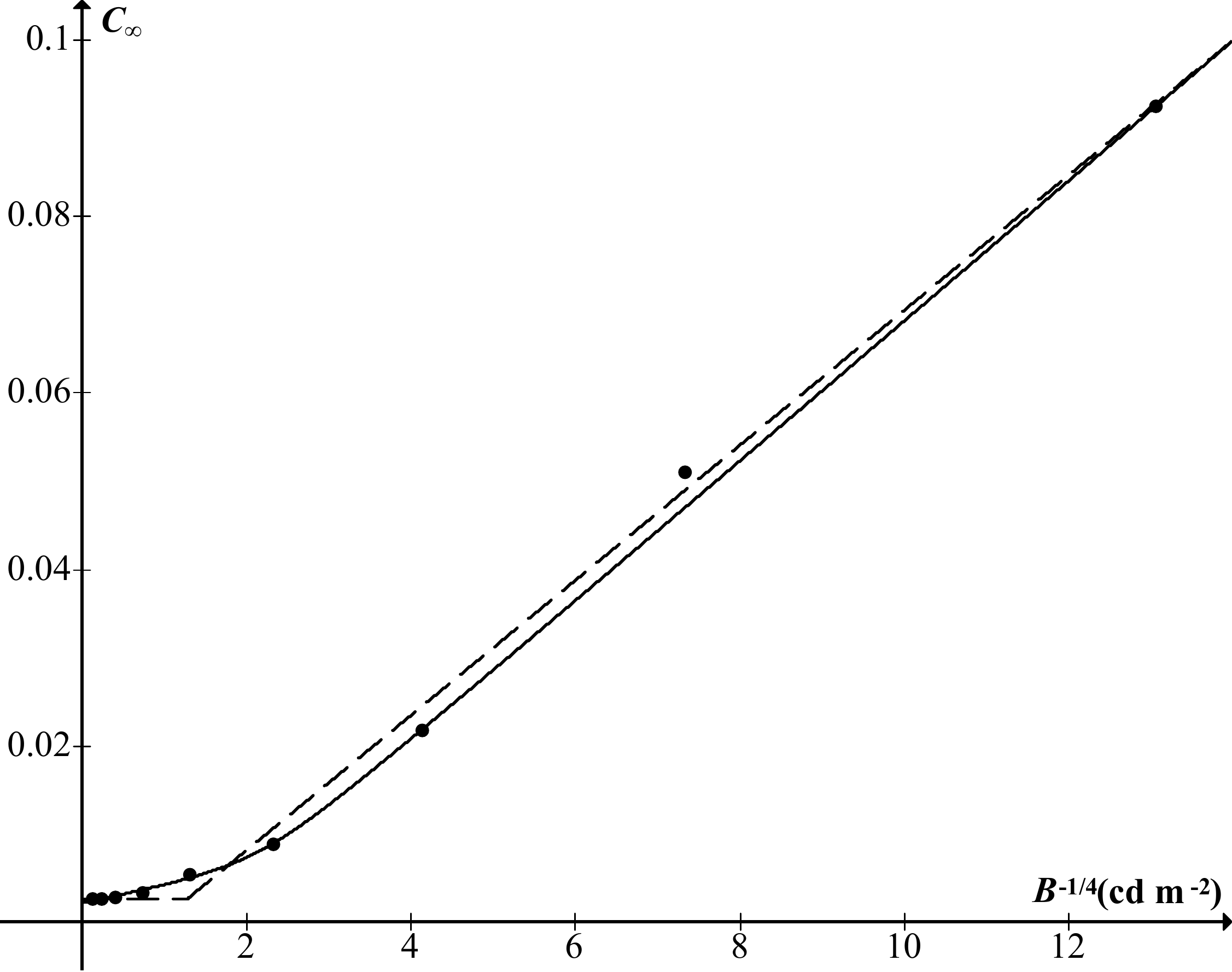}
\caption{Large-target contrast threshold, $C_{\mathrm{\infty}}$, as a function of luminance $B^{-1/4}$. Data from table 2 of \citet{taylorlarge2}. Dashed line is Eqs. \ref{kscot} to \ref{k3k4}; solid line is Eqs. \ref{khyp}, \ref{bi}.}
\label{cinf}
\end{figure}

In view of the uncertainty, \citet{taylorlarge2} offered upper and lower bounds for $C_{\mathrm{\infty}}$, subject to a number of assumptions. The upper bounds will be adopted here, since they are more compatible with Blackwell's figures, but with the understanding that the data are less robust than those used in the previous section for obtaining the function $R$. In fact, as shown in Fig. \ref{cinf}, the data display a similar luminance dependency to the one found for point sources, though now it is $C_{\mathrm{\infty}}$ rather than $\sqrt{R}$ that is plotted against $B^{-1/4}$, and the hyperbola has a more gradual bend, so that it is less accurate to regard it as consisting of two linear sections. The high-$B$ asymptote in this case is horizontal because of Weber's Law. Regression gives the coefficients as before (the data point for $B = 3.426\times10^{-4}$ cd m\textsuperscript{-2} being omitted to ensure convergence of $b_i$). This produces
\begin{equation}
C_{\mathrm{\infty}}^{\mathrm{scot}}= k_{1}B^{-1/4}+ k_{2},
\label{kscot}
\end{equation}
\begin{equation}
C_{\mathrm{\infty}}^{\mathrm{phot}}= k_{3}B^{-1/4}+ k_{4},
\label{kphot}
\end{equation}
\begin{equation}
k_{1}=7.633\times10^{-3}, k_{2}=-7.174\times10^{-3},
\label{k1k2}
\end{equation}
\begin{equation}
k_{3}=0, k_{4}=2.720\times10^{-3},
\label{k3k4}
\end{equation}
with split-point $B=3.54\times10^{-1}$ cd m\textsuperscript{-2}, and
\begin{equation}
C_{\mathrm{\infty}} = \surd(b_{1}B^{-1/2}+ b_{2}B^{-1/4}+ b_{3}) + b_{4}B^{-1/4}+ b_{5},
\label{khyp}
\end{equation}
\[
b_{1} = 9.606\times10^{-6}, b_{2} =-4.112\times10^{-5}, b_{3} =5.019\times10^{-5},
\]
\begin{equation}
b_{4} =4.837\times10^{-3}, b_{5} =-4.884\times10^{-3}. 
\label{bi}
\end{equation}

The same relation holds for the lower-bound series in \citet{taylorlarge2}, leading to slightly different coefficients. The full model is then constructed by smoothly joining the asymptotic sections $C_{\mathrm{low}}= R/A$ (with Eqs. \ref{rscot} and \ref{rphot}, or \ref{rhyp}) and $C_{\mathrm{high}}=C_{\mathrm{\infty}}$ (with Eqs. \ref{kscot} and \ref{kphot}, or \ref{khyp}):
\begin{equation}
C = ((R/A)^{q}+ C_{\mathrm{\infty}}^{q})^{1/q},
\label{fullmodel}
\end{equation} 
where $q$ is a parameter determined for best fit with the data. By construction, Eq. \ref{fullmodel} has the correct asymptotic behaviour for large and small $A$, with $q$ controlling the intermediate bend. At low light levels ($\mathrm{log}B \leq -0.5$) it is found that a constant value of $q$ is sufficient, however at higher levels one requires $q$ to be a function of luminance. Because of the discrete nature of the data it is not possible to specify an exact transition, but the following discontinuous function is found to be adequate:
\begin{equation}
q = 1.146 - 0.0885\mathrm{log}B, \quad B \geq 3.40 \mathrm {\,cd\, m^{-2}}
\label{qhigh}
\end{equation}
\begin{equation}
q = 0.8861 + 0.4\mathrm{log}B, \quad 0.193 \leq B < 3.40 \mathrm{\,cd\, m^{-2}}
\label{qmed}
\end{equation}
\begin{equation}
q = 0.6, \quad B < 0.193 \mathrm{\,cd\, m^{-2}}.
\label{qlow}
\end{equation}

\begin{figure}
\includegraphics[width=84mm]{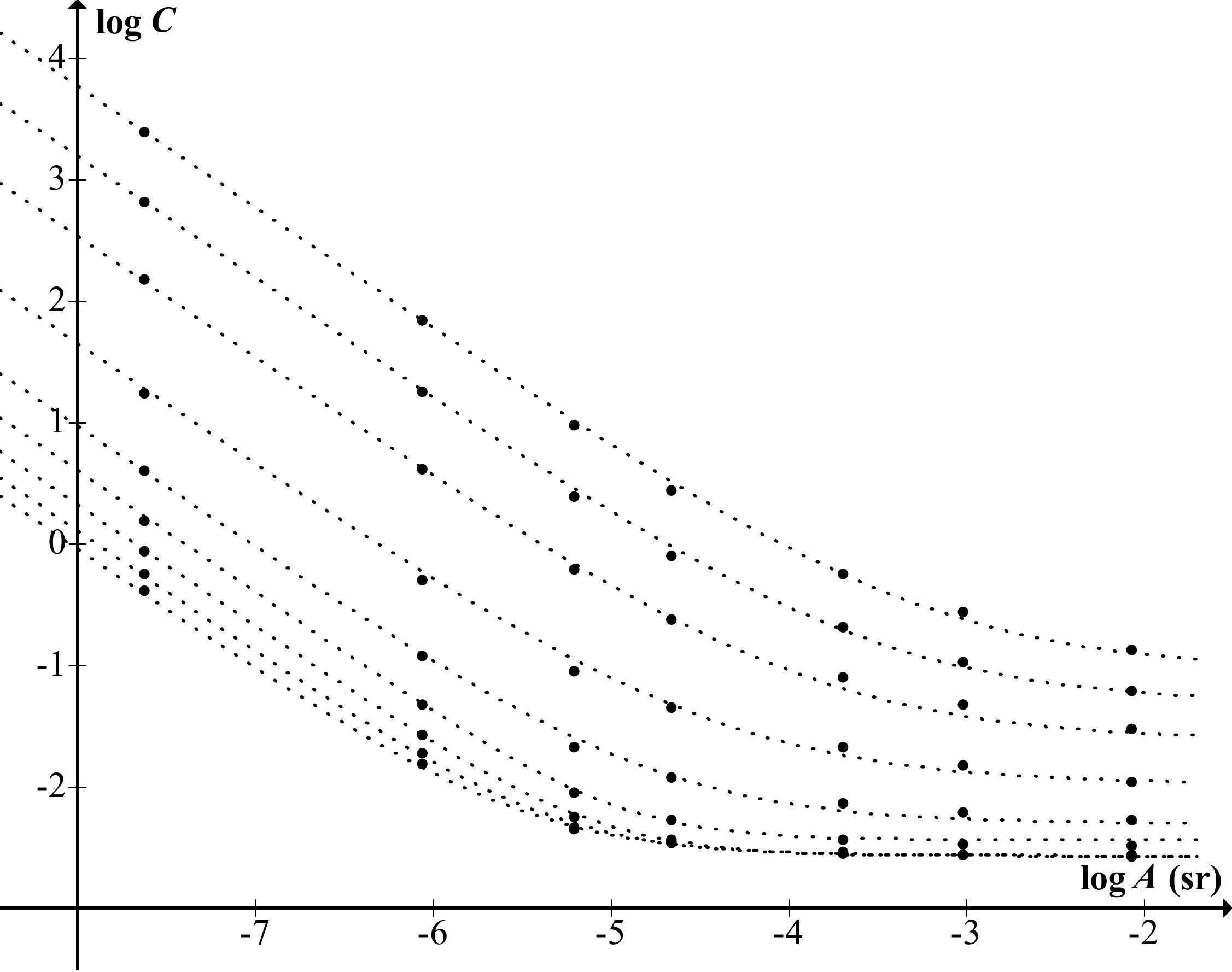}
\caption{Contrast threshold versus target size (Eq. \ref{fullmodel}) for luminances at intervals of one log-unit, $B = 3.426 \times 10^{-5}$ to $3.426 \times 10^{3}$ cd m\textsuperscript{-2} (top to bottom). Data from \citet{blackwell}. Top four curves modelled by Eqs. \ref{rscot}, \ref{r1r2}, \ref{kscot}, \ref{k1k2}, \ref{qlow}; middle two by Eqs. \ref{rhyp}, \ref{ai}, \ref{khyp}, \ref{bi}, \ref{qhigh}, \ref{qmed}; bottom three by Eqs. \ref{rphot}, \ref{r3r4}, \ref{kphot}, \ref{k3k4}, \ref{qhigh}.} 
\label{figall}
\end{figure}

The various forms of the model are shown in Fig. \ref{figall}. These show that at high or low light levels it is sufficient to use the simpler form of the model involving coefficients $r_i$ and $k_i$, while at intermediate (mesopic) levels one requires the more complicated model involving $a_i$ and $b_i$ coefficients. For astronomical visibility it is sufficient to use the scotopic model, whose upper limit of validity can be taken as approximately 0.1 cd m\textsuperscript{-2} (15 mag arcsec\textsuperscript{-2}) for achromatic sources. Consequently the other model forms will not be considered further in this article.

It has already been noted that the model becomes invalid at very low luminance, since $\Delta B = BC$ with $C$ given by Eq. \ref{fullmodel} does not tend to a non-zero limiting value $\Delta B_{0}$ as $B \rightarrow 0$. One can however join the existing model for $\Delta B$ to the zero-background asymptote $\Delta B_{0}$ with the same technique of geometric combination that has been used to join the small and large target-size asymptotes, i.e.
\begin{equation}
\Delta B_{\mathrm{full}} = (\Delta B_{0}^{n} + \Delta B^{n})^{1/n}.
\label{dblowest}
\end{equation}

\begin{figure}
\includegraphics[width=84mm]{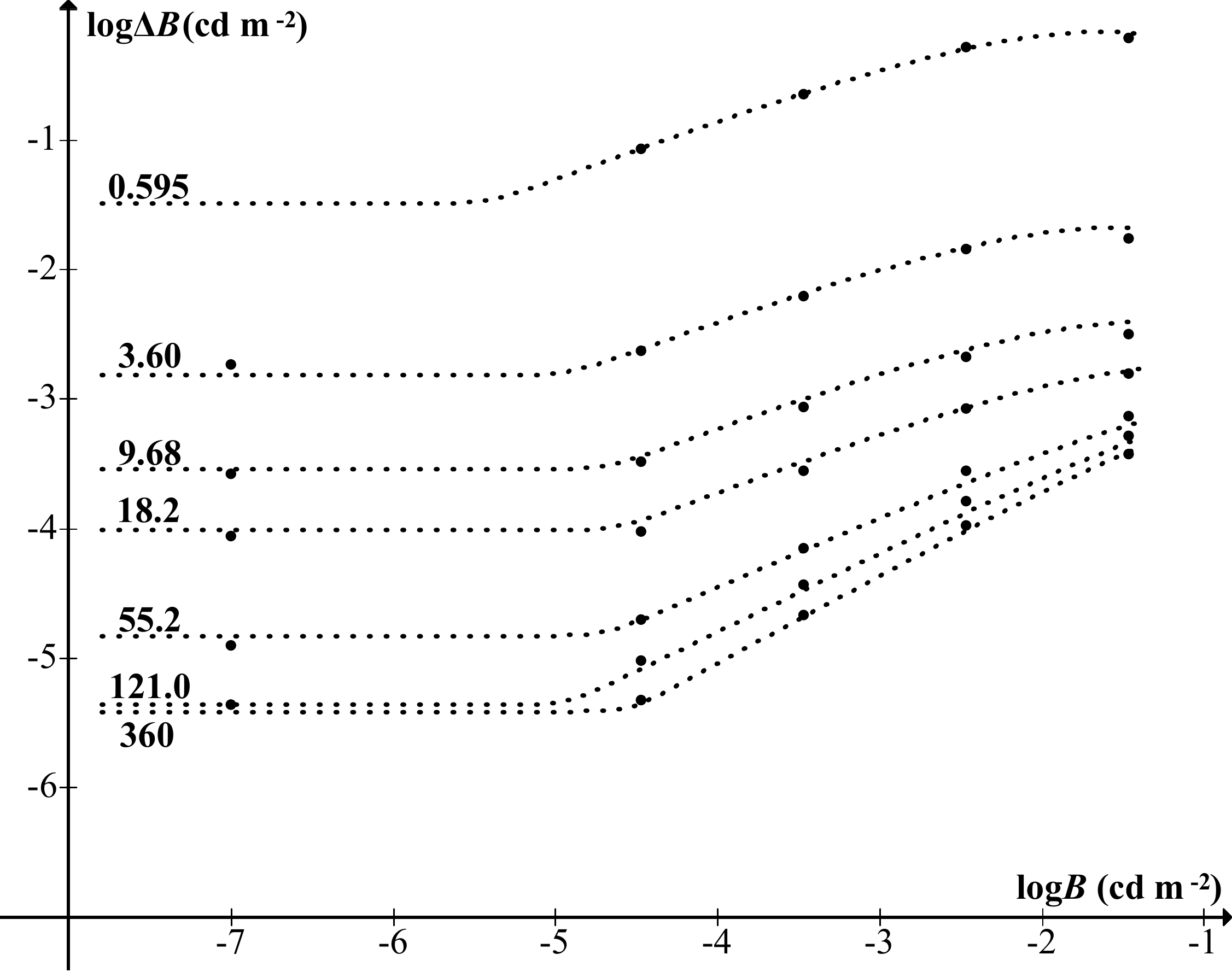}
\caption{Threshold increment $\Delta B$ at low luminance levels, with target diameters in arcmin. Data from \citet{blackwell} tables 4 and 8; dotted lines show Eq. \ref{lowmodel}.}
\label{figzero}
\end{figure}

To put this into effect one requires $\Delta B_{0}$ as a function of $A$. The data in table 4 of \citet{blackwell} show a linear relation yielding $\Delta B_{0} = 10^{-7.9591}A^{-0.8468}$, which tends to zero for large $A$. However the measurements are only for target diameters $3.6\--121$ arcmin, and one expects there to be a maximum $A$ beyond which there will be little or no further improvement. It can be estimated by recalling that for human vision the background becomes effectively zero at around $10^{-5}$ cd m\textsuperscript{-2} (Fig. \ref{deltabvb}), for which the Ricco area (Eq. \ref{riccoarea}) is $8.94 \times10^{-4}$ sr, or 116 arcmin diameter. Hence the lowest measured threshold for 121 arcmin ($\Delta B = 10^{-5.4162}$ cd m\textsuperscript{-2}) can reasonably be taken as limit, and incorporated through geometric combination (with some exponent $p$), to give $\Delta B_{0}$, which is then used in Eq. \ref{dblowest}. The choice of exponents is somewhat arbitrary because of the lack of data for $0 < B < 3.426 \times 10^{-5}$ cd m\textsuperscript{-2}, but the choice $n=9$, $p = 6$ proves adequate. Then (dropping the subscript `full')
\begin{eqnarray}
\Delta B
=[(10^{-47.7546}A^{-5.0808} + 10^{-32.4971})^{1.5}\nonumber\\
+B^{9}((r_{1}B^{-0.25}+r_{2})^{1.2}A^{-0.6}+(k_{1}B^{-0.25}+k_{2})^{0.6})^{15}]^{1/9},\nonumber\\
 \label{lowmodel}
\end{eqnarray}
which is shown in Fig \ref{figzero} (with the zero-background data plotted at $\mathrm{log}B=-7$). In fact this model version will not be considered further in this article, since the abrupt transition to an effectively zero background means that in practical applications sufficient accuracy can be achieved using the simpler model version with a cut-off at $10^{-5}$ cd m\textsuperscript{-2}. So the model will henceforth always be assumed to be
\begin{equation}
C = [((r_{1}B^{-1/4} + r_{2})^2/A)^{3/5} + (k_{1}B^{-1/4} + k_{2})^{3/5}]^{5/3},
\end{equation}
\begin{equation}
r_{1}=6.505\times10^{-4}, r_{2}=-8.461\times10^{-4},
\label{ri}
\end{equation}
\begin{equation}
k_{1}=7.633\times10^{-3}, k_{2}=-7.174\times10^{-3},
\label{ki}
\end{equation}
for $10^{-5} \leq B \leq 3.426 \times 10^{-2}$ cd m\textsuperscript{-2}, and
\begin{equation}
C = [(\xi_{1}/A)^{3/5} + \xi_{2})^{3/5}]^{5/3},
\end{equation}
\begin{equation}
\xi_{1} = (10^{5/4}r_{1}+ r_{2})^{2} = 1.150 \times 10^{-4},
\end{equation}
\begin{equation}
\xi_{2} = (10^{5/4}k_{1}+ k_{2}) = 1.286 \times 10^{-1},
\end{equation}
for $0 < B \leq 10^{-5}$ cd m\textsuperscript{-2}.

\section{Astronomical visibility}
\subsection{Naked-eye}
\label{nakedeye}

For naked-eye star visibility it is sufficient to use the $A \rightarrow 0$ limit of the threshold curve, and the natural brightness of the sky means that the zero-background limit is not required. Blackwell's backgrounds can be taken as sufficiently representative of the night sky without excessive light pollution. Then for scotopic vision the point-source formula Eq. \ref{deltaiscot} applies with Blackwell values, Eq. \ref{r1r2},

\begin{equation}
\Delta I = F(6.505\times10^{-4}B^{1/4} -8.461\times10^{-4}B^{1/2})^{2},
\label{deltai}
\end{equation}
where $\Delta I$ is the illuminance of the star in the absence of atmosphere (which contributes to the increment, as explained earlier), and the field factor $F$ has been introduced, assumed to include all factors associated with the target and medium, as well as laboratory scaling (for actual detection) and the personal factor of the observer. $B$ is for the area immediately surrounding the target, and it is assumed that the target remains visible long enough for scintillation to be excluded. In astronomical units (magnitude limit $m_0$, sky surface brightness $\mu_{\mathrm{sky}}$, zero-point $Z = 2.54 \times 10^{-6}$ lux), Eq. \ref{deltai} is well approximated by the linear functions
\begin{equation}
m_0=0.3834\mu_{\mathrm{sky}} - 1.4400 - 2.5\mathrm{log}F,
\label{approxnelm}
\end{equation}
if $20 < \mu_{\mathrm{sky}} < 22$ mag arcsec\textsuperscript{-2} (maximum error 0.01 mag),
\begin{equation}
m_0=0.4260\mu_{\mathrm{sky}} - 2.3650 - 2.5\mathrm{log}F,
\label{linapprox}
\end{equation}
if $21 < \mu_{\mathrm{sky}} < 25$ mag arcsec\textsuperscript{-2} (maximum error 0.04 mag).

\begin{figure}
\includegraphics[width=84mm]{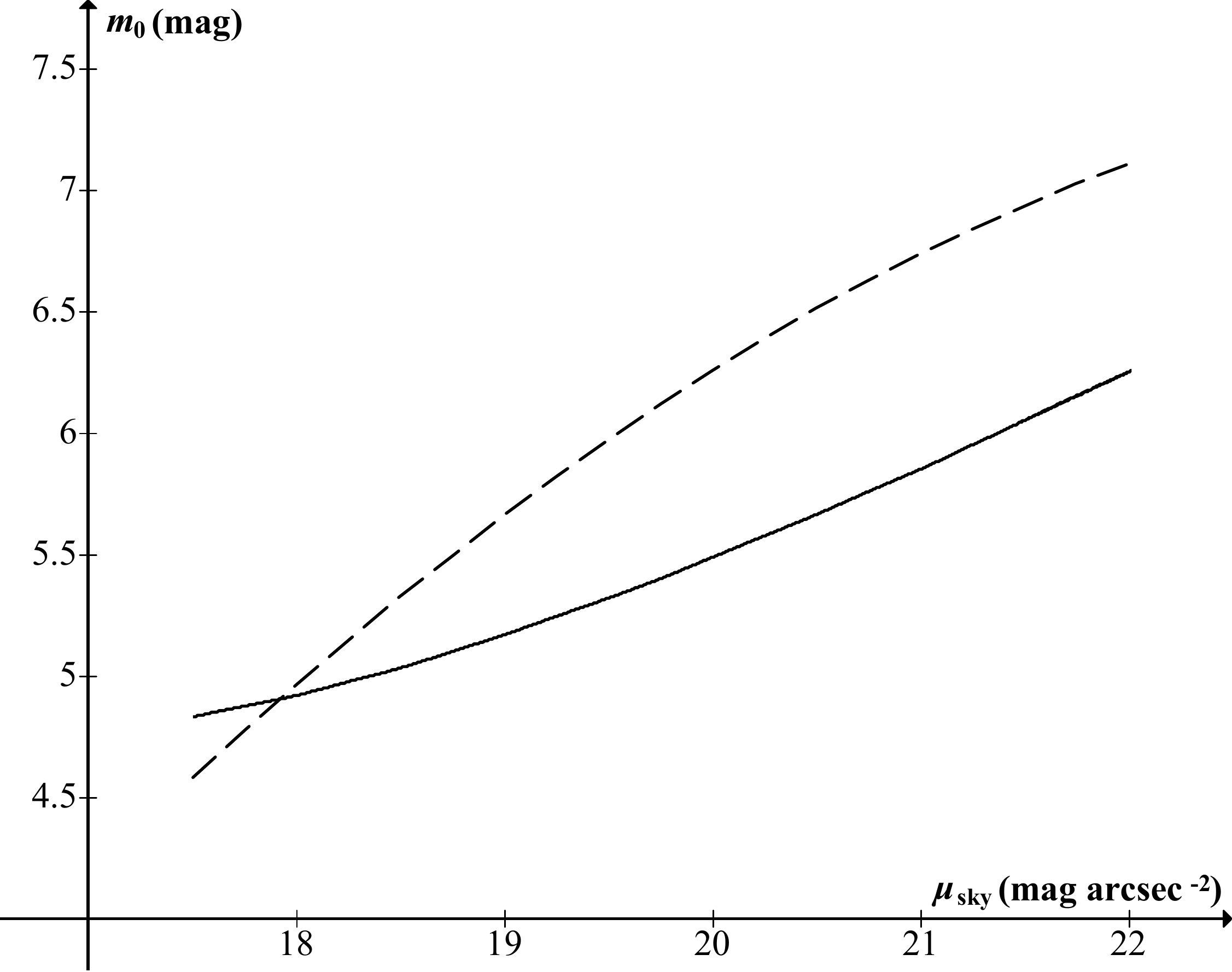}
\caption{Naked-eye limiting magnitude $m_0$ as a function of sky surface brightness $\mu_{\mathrm{sky}}$. Solid line is Eq. \ref{deltai} with $F=2$; dashed line is Hecht's model (Eq. \ref{hecht}) without rescaling.}
\label{mevhecht}
\end{figure}

For a dark sky with $B = 2 \times 10^{-4}$ cd m\textsuperscript{-2} (21.83 mag arcsec\textsuperscript{-2}) Eq. \ref{deltai} gives a magnitude limit $m_0 = 6.93 - 2.5\mathrm{log}F$. This would suggest that in actual observing situations $F$ is typically somewhere between 2.4 and 1.4 (giving limits 5.98 to 6.57 mag), with 7 mag corresponding to $F =0.94$. In view of the historical evidence discussed earlier, it would seem that for illustrative purposes a notional value $F =2$ (limit 6.18 mag) could be taken as a typical overall field factor. Fig. \ref{mevhecht} shows limiting magnitude as a function of sky surface brightness from Eq. \ref{deltai} with $F=2$. Also plotted is Eq. \ref{hecht} converted to astronomical units (without field factor rescaling). Either curve can be moved up or down by a choice of overall field factor; what is significant is the incorrect curvature of Hecht's formula remarked earlier (reversed now because of the change to astronomical units).

The $A \rightarrow \infty$ asymptote of the threshold curve, $C_{\mathrm{\infty}}$, gives the lower limit of visibility for large targets. From Eqs. \ref{c_def}, \ref{kscot} and \ref{k1k2}, the above-atmosphere luminance limit is
\begin{equation}
\Delta B_{\mathrm{\infty}} = F(7.633\times10^{-3}B^{3/4}-7.174\times10^{-3}B).
\label{sblim}
\end{equation}
In astronomical units this gives the limiting surface brightness $\mu_{\infty}$ for effectively infinite targets; with $\mu_{\mathrm{sky}} = 21.83$ mag arcsec\textsuperscript{-2} it is $\mu_{\infty}=24.94 - 2.5\mathrm{log}F$. Eq. \ref{sblim} is well approximated by
\begin{equation}
\mu_{\infty} = 0.6864\mu_{\mathrm{sky}} + 9.9325 - 2.5\mathrm{log}F,
\label{sblinear}
\end{equation}
if $18 < \mu_{\mathrm{sky}} < 22$ mag arcsec\textsuperscript{-2} (maximum error 0.02 mag arcsec\textsuperscript{-2}). For a general target of finite size the threshold increment (from Eqs. \ref{c_def}, \ref{fullmodel}, \ref{deltai}, \ref{sblim}) is
\begin{equation}
\Delta B = \left(\left(\frac{\Delta I}{A}\right)^{q}+ \Delta B_{\infty}^{q}\right)^{1/q},
\label{bcurve}
\end{equation}
and the Ricco area as conventionally defined (Eq. \ref{riccoarea} with Eqs. \ref{rscot}, \ref{kscot}, \ref{deltai} and \ref{sblim}) is
\begin{equation}
A_{\mathrm{R}} = \frac{(r_{1}B^{-1/4}+ r_{2})^{2}}{(k_{1}B^{-1/4}+ k_{2})} = \frac{\Delta I}{\Delta B_{\mathrm{\infty}}}.
\label{arlow}
\end{equation}
Hence Eq. \ref{bcurve} can be written in astronomical units as
\begin{equation}
\mu_{\mathrm{lim}} = \mu_{\infty} - \frac{2.5}{q}\mathrm{log}\left(\left(\frac{A_{\mathrm{R}}}{A}\right)^{q}+1\right),
\label{scurve}
\end{equation}
or, since the magnitude limit $m_0 = -2.5\mathrm{log}(A_{\mathrm{R}}\Delta B_{\mathrm{\infty}}/Z)$, 
\begin{equation}
\mu_{\mathrm{lim}} = m_0 - \frac{2.5}{q}\mathrm{log}\left(\frac{1}{\alpha^q} + \frac{1}{\alpha_{\mathrm{R}}^q}\right) + 5\mathrm{log}60,
\label{scurve2}
\end{equation}
for target and Ricco areas $\alpha$, $\alpha_{\mathrm{R}}$ in arcmin\textsuperscript{2} (and $\mu_{\mathrm{lim}}$ in mag arcsec\textsuperscript{-2}). By definition the target has magnitude $m_{\mathrm{lim}} = \mu_{\mathrm{lim}} - 2.5\mathrm{log}\alpha - 5\mathrm{log}60$, hence
\begin{equation}
m_{\mathrm{lim}} = m_0 - \frac{2.5}{q}\mathrm{log}\left(\left(\frac{\alpha}{\alpha_{\mathrm{R}}}\right)^{q}+1\right).
\label{mdash}
\end{equation}

The Ricco radius $r_\mathrm{R} = \sqrt{\alpha_{\mathrm{R}}/\pi}$ is well approximated by
\begin{equation}
r_\mathrm{R} = 5.21\mu_{\mathrm{sky}} - 76.2,
\label{radius}
\end{equation}
for $21 \leq \mu_{\mathrm{sky}} \leq 22$ mag arcsec\textsuperscript{-2} (maximum error 0.05 arcmin). This is considerably larger than the critical visual radius $r_{\mathrm{crit}}$ \citep{blackwell}, though either is only an approximation of the size at which an object becomes clearly extended \citep{nelson}. For $B = 2 \times 10^{-4}$ cd m\textsuperscript{-2} ($\mu_{\mathrm{sky}} = 21.83$ mag arcsec\textsuperscript{-2}), $r_{\mathrm{crit}}$ is approximately 4.5 arcmin (from figure 17 of \citet{blackwell}) while $r_\mathrm{R}=37.6$ arcmin. From Eq. \ref{scurve} the threshold surface brightness for a Ricco-area target is $-4.167\mathrm{log}(2)=1.25$ mag arcsec\textsuperscript{-2} brighter than $\mu_{\infty}$, and from Eq. \ref{mdash} the magnitude is likewise 1.25 mag brighter than $m_0$. This reflects the familiar fact that extended sources must be sufficiently brighter than the point-source limit in order to be seen as non-stellar, though the criterion is not stringent. As target size decreases, the magnitude threshold approaches $m_0$, while with increasing size the surface-brightness threshold approaches $\mu_{\infty}$, illustrating the fact that magnitude is a good visibility indicator for small targets, while surface brightness is better for large ones.

This can be applied to the visibility of M33. \citet{weaver} cited Lundmark's ability to see the galaxy without aid as an example of exceptional acuity, but at a Bortle Class 1 site it is an `obvious naked-eye object', and it is only in the fifth out of nine classes (`suburban sky') that M33 is considered undetectable \citep{bortle}. Tables 2 and 4 of \citet{devaucouleurs33} give the galaxy's total magnitude as 5.8, suggesting easy visibility at a dark site, but the foregoing remarks imply that a fainter stellar limit would be required. The data give an equivalent circular radius of 25.3 arcmin, but this is to an isophote 25.3 mag arcsec\textsuperscript{-2}, fainter than the eye can detect.

\begin{figure}
\includegraphics[width=84mm]{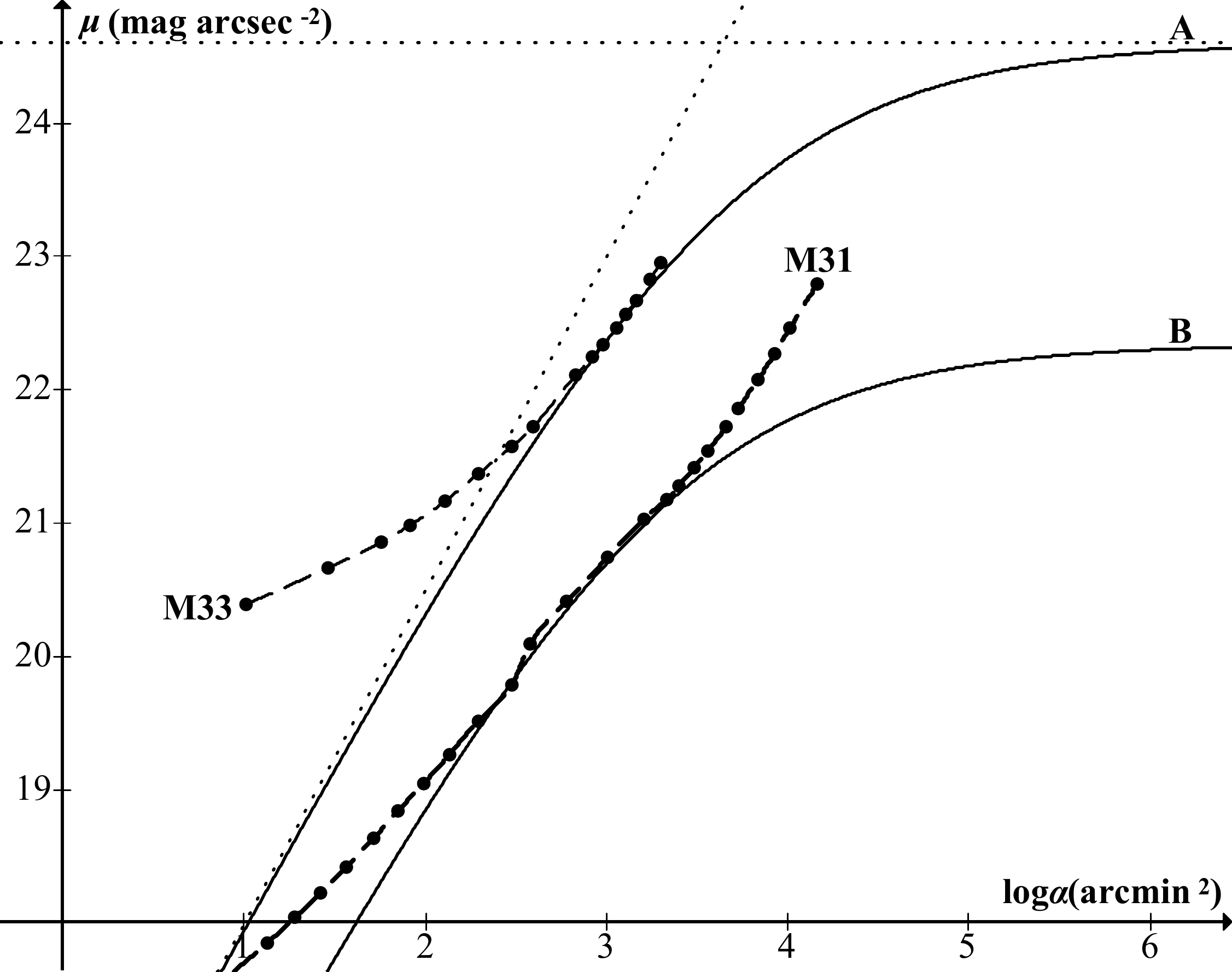}
\caption{Surface brightness profiles of M31 \citep{devaucouleurs31} and M33 \citep{devaucouleurs33}, with threshold curves A, B calculated from Eq. \ref{bcurve} such that each galaxy is just visible. Curve A asymptotes indicate the limiting magnitude and surface brightness for M33 to be just visible.}
\label{m33fig}
\end{figure}

Fig. \ref{m33fig} plots surface brightness versus log-area, so any line $\mu = 2.5\mathrm{log}\alpha + c$ is a line of constant magnitude $c - 5\mathrm{log}60$. Data points show the enclosed area and average surface brightness for successive isophotes of M33 (with an interpolated curve, dashed) as well as the threshold curve Eq. \ref{bcurve} (curve A, solid) for a background $\mu_{\mathrm{sky}}=21.83$ mag arcsec\textsuperscript{-2} and $F=1.378$, the latter parameter having been chosen so that the two curves are just touching, i.e. the target is at threshold. Raising $F$ will shift the threshold curve downwards, so the target becomes invisible. The co-ordinates of the intersection point give the visible size and brightness of the galaxy: equivalent circular radius 18.7 arcmin, surface brightness 22.43 mag arcsec\textsuperscript{-2}, magnitude 5.93. The Ricco asymptote is a line of constant magnitude 6.59 (the stellar limit required for the target to be visible) while the horizontal asymptote shows a large-target limit 24.59 mag arcsec\textsuperscript{-2}.

The same procedure can be repeated for different values of the surrounding sky background $\mu_{\mathrm{sky}}$ (the zenith value would generally be darker). Visibility at $F=2$ is found to require $\mu_{\mathrm{sky}}=22.63$ mag arcsec\textsuperscript{-2}, darker than the natural sky. For $21 \leq \mu_{\mathrm{sky}} \leq 22$ mag arcsec\textsuperscript{-2}, $F \approx 0.5482\mu_{\mathrm{sky}} - 10.585$ (maximum error 0.02). As $\mu_{\mathrm{sky}}$ increases within this range the meeting point of the curves moves very little (slowly upwards), so that for all $\mu_{\mathrm{sky}}$ the visible target has total magnitude 5.9 while the required stellar limit falls slowly from 6.67 to 6.57 mag. The visible radius and surface brightness hardly change as darkness increases within the stated range ($18.5\--18.75$ arcmin, $22.41\--22.44$ mag arcsec\textsuperscript{-2}); both differ substantially from the figures measured to the 25.3 mag arcsec\textsuperscript{-2} isophote, and instead refer to an (interpolated) isophotal limit $23.71\--23.75$ mag arcsec\textsuperscript{-2}. Some caution is necessary since the target is neither uniform nor circular, and the edge is actually seen against the (invisible) remainder of the galaxy rather than the sky, but the general conclusion (given the low $F$ values required) is that M33 can not reasonably be considered an easy target for average observers, even under very dark skies. Since the condition of its being just visible is a sustained limiting stellar magnitude of approximately 6.6 (certainly achievable under dark skies by observers with above-average acuity), the required magnitude limit can be taken as a sufficient sky quality indicator. That figure, which can be thought of as the galaxy's `effective' visual magnitude, is consistent with the visual estimate of approximately 7 mag made by \citet{holetschek} and accepted by \citet{hubblenebulae}. As noted earlier, the galaxy was included by \citet{heis} in his naked-eye star atlas which had stellar limit 6.7 mag. \citet{weaver} gave the galaxy's visual magnitude as 6.8 mag.

A similar procedure can be applied to M31 using data from \citet{devaucouleurs31}. It is found that with $F=2$ the galaxy should become just visible at approximately $\mu_{\mathrm{sky}}=19.2$ mag arcsec\textsuperscript{-2} (curve B on Fig. \ref{m33fig}), with visible area approximately 2100 arcmin\textsuperscript{2} and effective visual magnitude 5.2. This must be treated with caution since the luminance is close to mesopic, however the general prediction is that M31 should be an easy naked-eye target for average observers under moderately dark conditions, which accords with experience.

One should also consider the $B-V$ colour indices of M31 and M33, given by De Vaucouleurs as 0.91 and 0.55. \citet{cinzano2001} took the typical colour index of naked-eye stars as 0.7, and if this is considered the standard by which visual threshold is assessed then (from Eq. \ref{starcorr}) the effective magnitude of M31 should be lowered by 0.06 while that of M33 should be raised by 0.04. If colour index 0 is the standard then the effective magnitudes of M31 and M33 are instead lowered by 0.25 and 0.15.

\subsection{Point-source telescopic visibility}

For stars seen through a telescope, various authors \citep{garstang2000} proposed that the limiting magnitude would be given by
\begin{equation}
 m = N + 5\mathrm{log}D,
\label{classical}
\end{equation}
where $D$ is the entrance pupil diameter and $N$ is a constant. If $D$ is in centimetres then values of $N$ proposed in the literature cited by Garstang range from 6.8 to 8.7. In general, however, one must also take account of the background luminance, the magnification $M$ (or exit pupil diameter $d = D/M$), and field factors, in which case $N$ would need to be replaced by a function of these. \citet{schaefer} did this using Hecht's formula (Eq. \ref{hecht}). The same will now be done using the new model, giving results which can be tested against existing data.

Assume the same conditions under which Eq. \ref{deltai} applies. A star at threshold in a telescope will have apparent illuminance at the eye
\begin{equation}
\Delta I_{\mathrm{a}}= F_{\mathrm{T}}F_{\mathrm{M}}F(r_{1}B_{a}^{1/4}+ r_{2}B_{a}^{1/2})^{2}
\label{tel}
\end{equation}
where $B_{\mathrm{a}}$ is the apparent sky luminance in the eyepiece (the natural sky background darkened by magnification and light loss in the telescope), and $F_{\mathrm{T}}$ and $F_{\mathrm{M}}$ are field factors associated with telescope use, with $F_{\mathrm{T}}$ being the product of magnification-independent factors and $F_{\mathrm{M}}$ the product of magnification-dependent ones. As discussed in Section \ref{telescopeuse}, it should generally be sufficient to assume $F_{\mathrm{T}} = \sqrt{2}$ and $F_{\mathrm{M}} = 1$. Eq. \ref{tel} becomes invalid if magnification renders star images no longer point-like, or darkens the sky below about $10^{-5}$ cd m\textsuperscript{-2}. The latter effect will be incorporated by imposing a zero-background cut-off.

Let $\Delta I$ be the increment illuminance at the entrance pupil of a star at threshold in the eyepiece, and define $\delta_{\mathrm{min}}= \mathrm{min}(d,p)$, $\delta_{\mathrm{max}}= \mathrm{max}(d,p)$, for exit and eye pupil diameters $d$ and $p$. Let $F_\mathrm{t}^{-1}$ be the telescope's transmittance. Then, following \citet{tousey},
\begin{equation}
B_{\mathrm{a}}= \left(\frac{\delta_{\mathrm{min}}}{p}\right)^{2}\frac{B}{F_{\mathrm{t}}},
\label{ba}
\end{equation}
\begin{equation}
\Delta I_{\mathrm{a}}= \left(\frac{D}{\delta_{\mathrm{max}}}\right)^{2}\frac{\Delta I}{F_{\mathrm{t}}},
\end{equation}
hence (from Eq. \ref{tel})
\begin{equation}
\Delta I = \left(\frac{\delta_{\mathrm{max}}}{D}\right)^{2}F_{\mathrm{t}} F_{\mathrm{M}}F_{\mathrm{T}}F (r_{1}B_{\mathrm{a}}{}^{1/4}+ r_{2}B_{\mathrm{a}}{}^{1/2} )^{2} .
\label{telescope_limit}
\end{equation}

\begin{figure}
\includegraphics[width=84mm]{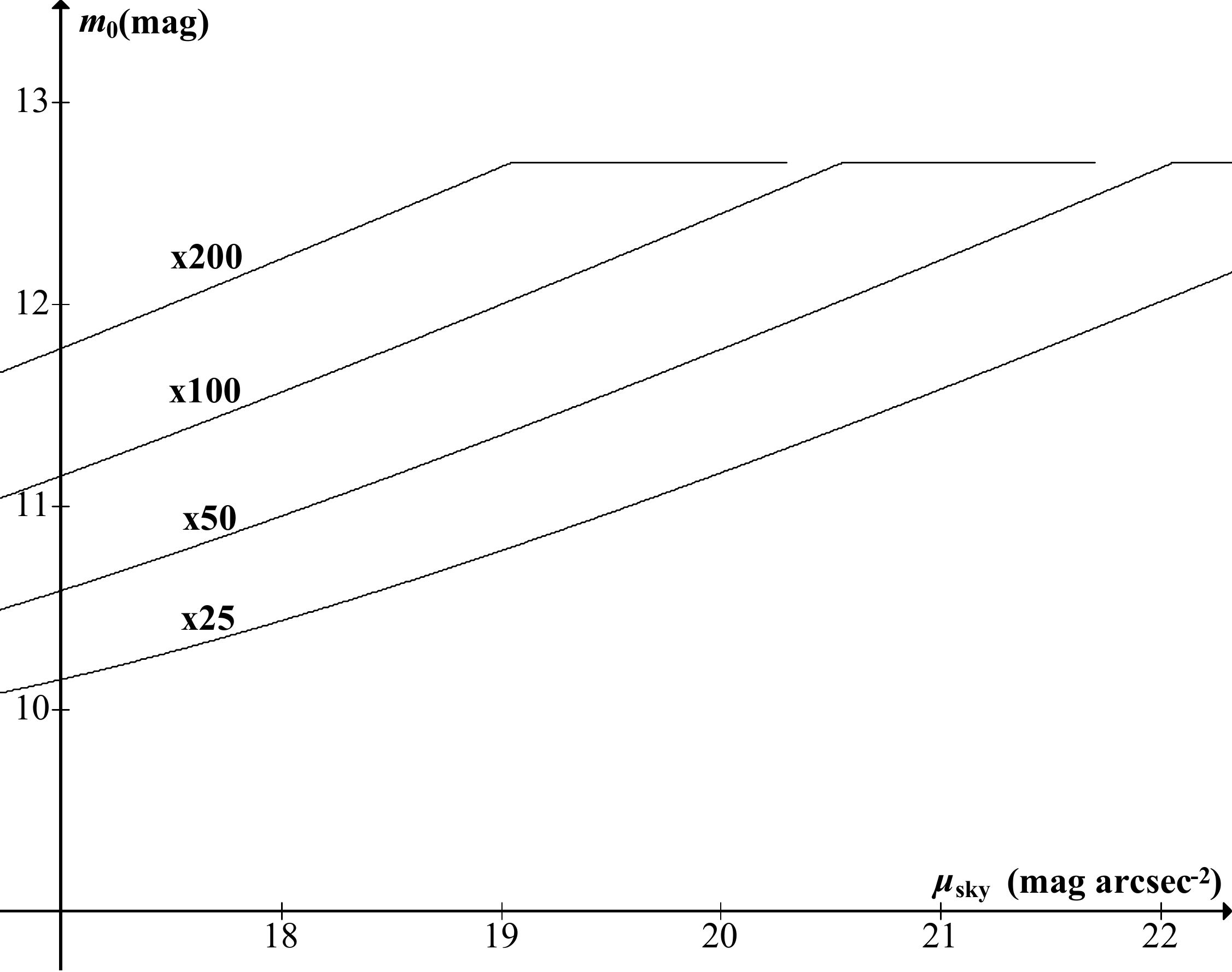}
\caption{Magnitude limit $m_0$ as a function of sky brightness $\mu_\mathrm{sky}$, calculated from Eq. \ref{telescope_limit} (with $F_{\mathrm{t}}F_{\mathrm{M}}F_{\mathrm{T}}F = 3.77$,  $p=7$ mm) for a telescope with clear aperture 100mm at various magnifications. The cut-off $m_\mathrm{cut} = 12.7$ mag is due to the background in the eyepiece becoming effectively zero.}
\label{tel_lim}
\end{figure}

Using Eq. \ref{linapprox} a very good approximation is found to be 
\begin{eqnarray}
m_0 = 0.426\mu_{\mathrm{sky}} - 2.365 + 5\mathrm{log}(D/\delta_{\mathrm{max}}) - 2.131\mathrm{log}(\delta_{\mathrm{min}}/p) \nonumber\\
 - 1.435\mathrm{log}F_{\mathrm{t}}  - 2.5\mathrm{log}(F_{\mathrm{M}}F_{\mathrm{T}}F),
\label{approxmag}
\end{eqnarray}
which gives the limit $m_0$ at magnification $M = D/p$. The threshold is assumed constant for $B_{\mathrm{a}} \leq 10^{-5}$ cd m\textsuperscript{-2}, which occurs for exit pupil $d \leq d_{0}$ where
\begin{equation}
d_{0} = p\sqrt\frac{10^{-5}F_{\mathrm{t}}}{B},
\label{d0}
\end{equation}
the cut-off threshold being
\begin{equation}
\Delta I_{\mathrm{cut}} =  \zeta \left(\frac{p}{D}\right)^{2}F_{\mathrm{t}}F_{\mathrm{M}}F_{\mathrm{T}}F,
\label{deltai0}
\end{equation}
where $\zeta = (10^{-5/4}r_{1}+ 10^{-5/2}r_{2})^{2} = 1.150 \times 10^{-9}$ lx. Hence the limiting magnitude for the telescope is
\begin{equation}
m_\mathrm{cut} = 5\mathrm{log}D - 2.5\mathrm{log}(Z^{-1}\zeta p^{2}F_{\mathrm{t}}F_{\mathrm{M}}F_{\mathrm{T}}F),
\label{m0}
\end{equation}
where $Z = 2.54 \times 10^{-6}$ lx. Taking $F_{\mathrm{M}} = 1$, $F_{\mathrm{T}} = \sqrt{2}$ and typical values $p = 7 \times 10^{-3}$ m, $F_{\mathrm{t}} = 1.33$ (75 per cent transmittance) gives
\begin{equation}
m_\mathrm{cut} = 5\mathrm{log}D + 8.45 - 2.5\mathrm{log}F,
\end{equation}
for $D$ in cm. This is to be compared with Eq. \ref{classical}. The proposed range of values for $N$ would correspond to $F$ ranging $0.79 \-- 4.55$, while with the notional value $F = 2$ we obtain $N = 7.69$, which agrees with Sinnott's figure 7.7 cited by \citet{garstang2000} as the best value for general use. Fig. \ref{tel_lim} shows the limit as a function of sky surface brightness for a telescope with entrance pupil 0.1m.

Eqs. \ref{approxmag} and \ref{m0} imply that the graph of $m_0$ versus $-\mathrm{log}d$ consists of three straight sections with gradients 5 ($d \geq p$), 2.131 ($p \geq d \geq d_{0}$) and 0 ($d \leq d_{0}$). This can be tested against the data of \citet{bowen} who recorded his threshold for various exit pupils using refractors of aperture 0.33 inch ($8.38 \times10^{-3}$ m), 6 inch ($1.52 \times10^{-1}$ m), and the 60-inch (1.52m) reflector at Mount Wilson, with entrance pupil diameter $D=1.39$ m \citep{seares}. The most extensive results were for the 6-inch telescope, with data falling clearly into three sections fitted by
\begin{eqnarray}
m_0 &=& -5\mathrm{log}d + 1.02, \nonumber\\
m_0 &=& -2.131\mathrm{log}d + 7.57, \nonumber\\
m_0 &=& 13.96.
\label{6inchlines}
\end{eqnarray}
This is consistent with targets having been effectively equivalent (stars of roughly equal colour index and zenith angle), observed under effectively uniform conditions, so that $F$ can be regarded as constant. The intersection of the first two lines gives Bowen's pupil diameter as 5.2mm (consistent with his age, 49 years) while that of the second pair fixes $d_{0} = 1.0$ mm implying $B/F_{\mathrm{t}} = 2.70 \times 10^{-4}$ cd m\textsuperscript{-2} (from Eq. \ref{d0}). Eq. \ref{telescope_limit} then gives $F_{\mathrm{t}}F_{\mathrm{M}}F_{\mathrm{T}}F = 4.78$.

\begin{figure}
\includegraphics[width=84mm]{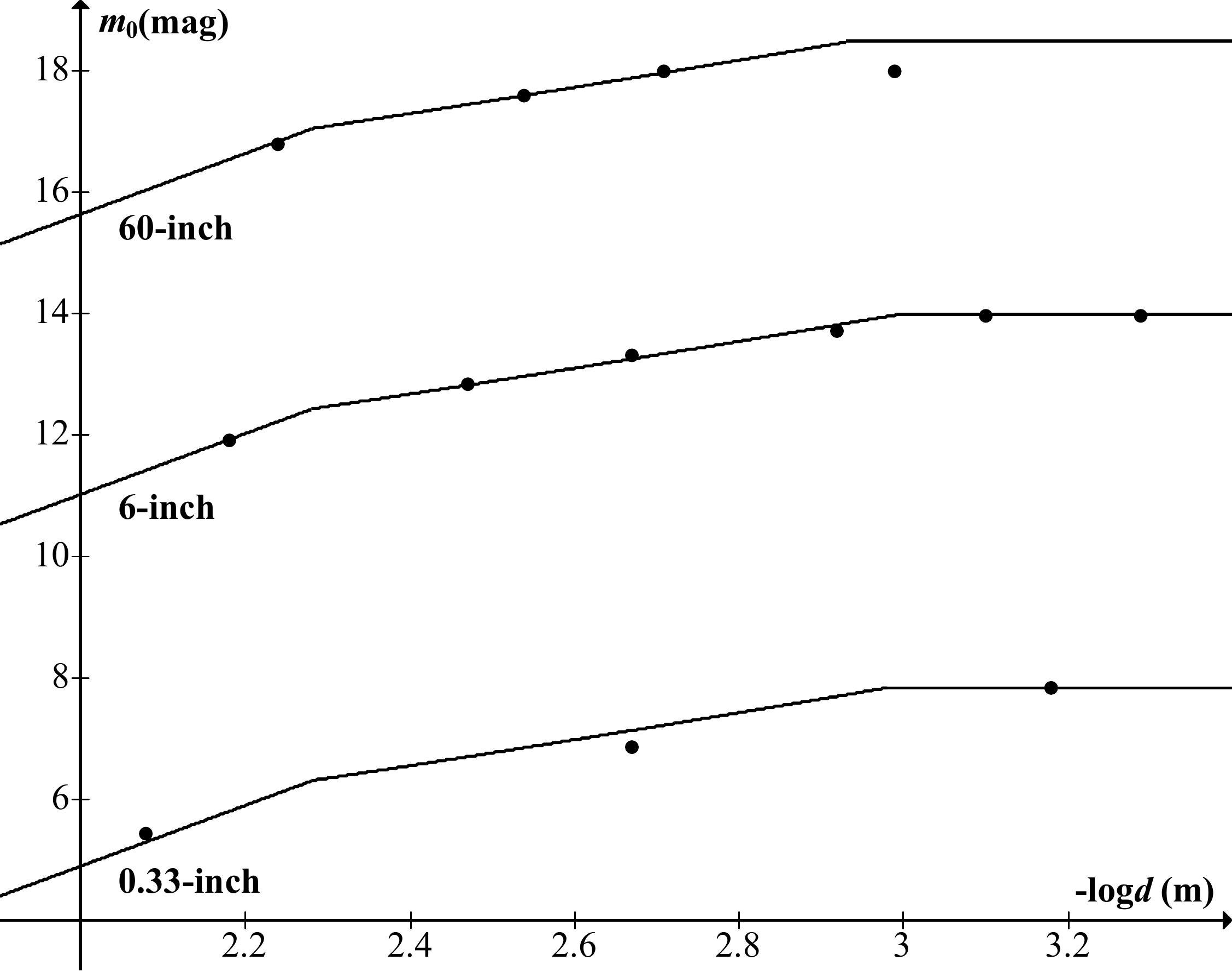}
\caption{Magnitude limit $m_0$ as a function of exit pupil $d$ for three telescopes (labelled by aperture). Data from \citet{bowen}, modelled with Eqs. \ref{telescope_limit}, \ref{deltai0}. The anomalous data point for the 60-inch at highest magnification, attributable to the stellar disc being no longer point-like, is modelled in Section \ref{telescopicthresholdcurve}.}
\label{bowen}
\end{figure}

The 60-inch data imply $p=5.0$ mm, consistent with the 6-inch figure (which is retained), but the highest magnifications seem to suggest a zero-background cut-off of 18.0 mag, which by Eq. \ref{d0} would produce an unreasonably low transmittance for any reasonable value of $B$. In fact Bowen considered the limit for highest magnification to be suspect due to poor seeing: the stellar discs would have had a diameter of more than 10 arcmin (the limit of validity of the point-source model) and Bowen found them `noticeably fuzzy'. This will be returned to once the general model for finite target sizes has been presented. Meanwhile, the more reliable $p \geq d \geq d_{0}$ line for the 60-inch telescope is
\begin{equation}
m_0 = -2.131\mathrm{log}d + 12.19
\label{60midline}
\end{equation}
which with Eqs. \ref{approxmag} and \ref{6inchlines} (assuming $F_{\mathrm{M}}F_{\mathrm{T}}$ to be uniform for all telescopes) yields
\begin{equation}
1.435\mathrm{log} \left( \frac{F_{\mathrm{t}}^{\mathrm{60in}}}{F_{\mathrm{t}}^{\mathrm{6in}}} \right) = 7.57-12.19+5\mathrm{log} \left(\frac{D^{\mathrm{60in}}}{D^{\mathrm{6in}}} \right),
\end{equation}
hence $F_{\mathrm{t}}^{\mathrm{60in}}= 1.35F_{\mathrm{t}}^{\mathrm{6in}}$. For the 0.33-inch telescope there are only three magnitude measurements, the lowest pair producing an anomalous and improbably high pupil diameter 7.3 mm, suggesting inaccuracy in the data. Using Eq. \ref{m0} with the highest measurement one finds  $F_{\mathrm{t}}^{\mathrm{0.33in}}= 0.90F_{\mathrm{t}}^{\mathrm{6in}}$. The predicted thresholds for all three telescopes can then be plotted using Eqs.  \ref{telescope_limit} and \ref{deltai0}, shown in Fig. \ref{bowen} to give very good agreement with the data. Table 1 of \citet{schaefer} contains predictions for all except the lowest 0.33-inch limit; when the dubious highest 60-inch limit is also excluded Schaefer's model has r.m.s. error 0.37 mag, compared with 0.09 mag for the present model.

The minimum possible value for $F_{\mathrm{t}}$ would have been 1.04 (1 per cent reflectance at four coated glass-air surfaces for a refractor with no light scattering). From the $B/F_{\mathrm{t}}$ value for the telescope with lowest $F_{\mathrm{t}}$ (0.33-inch) this produces a lowest bound $B = 3.12 \times 10^{-4}$ cd m\textsuperscript{-2} (21.35 mag arcsec\textsuperscript{-2}). Realistically estimating $90 \pm 5$ per cent transmitance for the 0.33-inch telescope, the previously calculated ratios give transmittances $81 \pm 4.5$ per cent for the 6-inch  and $60 \pm 3.3$ per cent for the 60-inch telescopes, the latter implying a reflectance of approximately 85 per cent at each of the three aluminized mirrors, a plausible figure not far below the maximum value of 89 per cent. The figure for the 6-inch refractor could suggest it was in need of cleaning, or did not have anti-reflection coating as Bowen stated (having the equivalent of 95 per cent transmittance at each air-glass surface), or perhaps the clear aperture was actually slightly less than the assumed value.

The figures imply a sky brightness $B = 3.34 \pm 0.19 \times 10^{-4}$ cd m\textsuperscript{-2} ($\mu_\mathrm{sky}= 21.27 \pm 0.06$ mag arcsec\textsuperscript{-2}). Assuming $F_{\mathrm{M}}F_{\mathrm{T}} = \sqrt{2}$ as usual, $F$ is then $2.74 \pm 0.15$, somewhat higher than the `typical' value 2 but consistent with Bowen's age, and giving his naked-eye limit (from Eq. \ref{deltai}) as $5.62 \pm 0.04$ mag. If he had recorded his naked-eye limit then the transmittances and sky brightness could have been determined from that.

\citet{garstang2000} analysed Bowen's data using a modified form of the Hecht equation, employing the enumerated field factor treatment of \citet{schaefer}. He made estimates of various parameters in his model, and arrived at a predicted sky brightness $B = 1.05 \times 10^{-3}$ cd m\textsuperscript{-2} ($\mu_\mathrm{sky} = 20.03$ mag arcsec\textsuperscript{-2}), substantially brighter than the new estimate. Garstang made an airmass correction based on a guess of the zenith angle of observed stars, to arrive at a zenith brightness $B = 8.59 \times 10^{-4}$ cd m\textsuperscript{-2} ($\mu_\mathrm{sky} = 20.24$ mag arcsec\textsuperscript{-2}). When the same correction is made to the new figure one obtains $B = 2.45 \times 10^{-4}$ cd m\textsuperscript{-2} ($\mu_\mathrm{sky} = 21.61$ mag arcsec\textsuperscript{-2}).

\citet{garstang2004} calculated the sky brightness at Mount Wilson throughout the twentieth century using his light pollution model, for which a crucial parameter is the average light emission per head of population. Garstang estimated this parameter, guided partly by his analysis of Bowen's data, producing results somewhat darker than his previous work (20.82 mag arcsec\textsuperscript{-2} for 1950), but still considerably brighter than the new estimate. The present finding suggests that light emissions from Los Angeles in the first half of the twentieth century were lower than Garstang assumed, and that the growth of light pollution in the second half was far more rapid than he calculated.

\subsection{The telescopic threshold curve}
\label{telescopicthresholdcurve}

The threshold for objects of angular area $A$, seen against a sky of luminance $B$, through a telescope with magnification $M$, is calculated by transforming Eq. \ref{fullmodel} to take account of the change of target size and background luminance imposed by the instrument, as well as the field factors discussed previously:
\begin{equation}
C = \phi\left(\left(\frac{R_{\mathrm{a}}}{A_{\mathrm{a}}}\right)^{q} + C_{\mathrm{a}}^{q}\right)^{1/q},
\label{ctel}
\end{equation}
where $\phi \equiv F_{\mathrm{T}}F_{\mathrm{M}}F$ and
\begin{equation}
R_{\mathrm{a}} = \left(\frac{r_{1}}{B_{\mathrm{a}}^{1/4}} + r_{2}\right)^{2},
\label{ra}
\end{equation}
\begin{equation}
A_{\mathrm{a}} = M^{2} A,
\end{equation}
\begin{equation}
C_{\mathrm{a}} = \frac{k_{1}}{B_{\mathrm{a}}^{1/4}} +k_{2},
\end{equation}
with $r_i$, $k_i$, $q$ and $B_{\mathrm{a}}$ given by Eqs. \ref{r1r2}, \ref{k1k2}, \ref{qlow} and \ref{ba}. The telescopic threshold curve has the same general shape as the naked-eye one, but with shifted asymptotes. The `telescopic Ricco area', $A_{\mathrm{TR}}$, can be defined as the area on the sky of a target whose image in the eyepiece has Ricco area with respect to the apparent background, i.e.
\begin{equation}
A_{\mathrm{TR}} = \frac{R_{\mathrm{a}}}{M^{2}C_{\mathrm{a}}}.
\end{equation}
Then Eq. \ref{ctel} can be rewritten in terms of the large-target and point-source limits, $\Delta B_{\infty} = \phi BC_{\mathrm{a}}$, $\Delta I = \phi BR_{\mathrm{a}}/M^{2}$, 
\begin{equation}
\Delta B = \Delta B_{\infty}\left(\left(\frac{A_{\mathrm{TR}}}{A}\right)^{q}+1\right)^{1/q},
\end{equation}
or
\begin{equation}
\Delta B = \Delta I\left(\frac{1}{A^{q}} + \frac{1}{A_{\mathrm{TR}}^{q}}\right)^{1/q},
\end{equation}
which in astronomical units ($\alpha$, $\alpha_{\mathrm{TR}}$ in arcmin, $\mu_\mathrm{lim}$ in mag arcsec\textsuperscript{-2}) give the telescopic equivalents of Eqs. \ref{scurve}, \ref{scurve2}:
\begin{equation}
\mu_\mathrm{lim} = \mu_{\infty} - \frac{2.5}{q}\mathrm{log}\left(\left(\frac{\alpha_{\mathrm{TR}}}{\alpha}\right)^{q}+1\right),
\label{scurvetel}
\end{equation}
\begin{equation}
\mu_\mathrm{lim} = m_0 - \frac{2.5}{q}\mathrm{log}\left(\frac{1}{\alpha^{q}} + \frac{1}{\alpha_{\mathrm{TR}}^{q}}\right)^{1/q} + 5\mathrm{log}60,
\label{scurvetel2}
\end{equation}
\begin{equation}
\mu_{\infty} = \mu_\mathrm{sky} - 2.5\mathrm{log}(\phi C_{\mathrm{a}}),
\label{sinf}
\end{equation}
\begin{equation}
m_0 = \mu_\mathrm{sky} + 2.5\mathrm{log}\left(\frac{\pi^2}{60^{4}180^{2}}.\frac{M^{2}}{\phi R_{\mathrm{a}}B}\right).
\label{m}
\end{equation}
Thus a threshold target of area $\alpha$ has magnitude
\begin{equation}
m_\mathrm{lim} = m_0 - \frac{2.5}{q}\mathrm{log}\left(\left(\frac{\alpha}{\alpha_{\mathrm{TR}}}\right)^{q}+1\right).
\label{magdiff}
\end{equation}

The zero-background cut-off imposed at $B_{\mathrm{a}}$ = $10^{-5}$ cd m\textsuperscript{-2} corresponds to an eyepiece Ricco area $R_{\mathrm{a}}/C_{\mathrm{a}}$ = $(10^{5/4}r_{1} + r_{2})^{2}/(10^{5/4}k_{1} + k_{2})$ = $8.941 \times 10^{-4}$ sr or 10,567 arcmin\textsuperscript{2}. Then the zero-background threshold is Eq. \ref{scurvetel} or \ref{scurvetel2} with $\alpha_{\mathrm{TR0}} = 10567/M_{0}^2$ arcmin\textsuperscript{2}, where (from Eq. \ref{d0}) $M_{0}^{2} = 10^{5}BD^{2}/(p^{2}F_{\mathrm{t}})$.

\citet{bowen} obtained a limit 18 mag for stars seen with the 60-inch telescope at $M=1500$, rather than the predicted point-source limit $18.7$ mag. From Fig. \ref{bowen} it can be seen that this was against an effectively zero background. With the parameters derived earlier, one finds $\alpha_{\mathrm{TR0}} = 7.012 \times 10^{-3}$ arcmin\textsuperscript{2}, and Eq. \ref{magdiff} can be solved for $\alpha$ (with $m_0-m_\mathrm{lim} = 0.7$). If this is interpreted as the area of the Gaussian stellar disc (with due caution regarding the target's non-uniformity) then it gives the diameter as 3.0 arcsec, and from Section \ref{turbulence} the FWHM seeing is estimated as 3.0/2.8 = 1.1 arcsec, entirely consistent with Bowen's remark that it was `about average'.

The model can be applied to the observations of William Herschel, who compiled three catalogues of `nebulae' (mostly galaxies) discovered between 1783 and 1802 with a telescope which had an 18.7-inch (475.0mm) diameter speculum mirror and `sweeping power' $M = 157$ \citep[v.1, p.260]{herschel}. From 1786 he used the telescope in `front-view' mode, without a secondary mirror \citep[v.1, p.xlii]{herschel} so that the entrance pupil was equal to the full aperture (assuming his head did not intrude) and the exit pupil diameter was $d = 3.03$ mm. In 1801, by looking at the star Vega through artificial pupils of various sizes, he measured his eye pupil as 0.2 inches (5.08mm)  \citep[v.2, p.585]{herschel}. He visually measured the reflectance of his mirror as 67 per cent and determined the overall transmittance (in front-view mode with a single-element eyepiece) as 63.8 per cent \citep[v.2, p.40]{herschel}, very close to the modern theoretical figure $0.68 \times 0.96^{2} = 62.7$ per cent, which gives $F_{\mathrm{t}} = 1.6$ for his telescope at best performance. The sky brightness would have varied during the observing period due to solar activity, but sunspot data are sparse \citep{zolotova}. The figure $B = 2 \times 10^{-4}$ cd m\textsuperscript{-2} (21.83 mag arcsec\textsuperscript{-2}) will be taken as an approximation. From Eq. \ref{d0} this gives $d_{0} = 1.45$ mm with $M=328$.

If it is assumed as before that $\phi = \sqrt{2}F$ where $F$ is the naked-eye field factor, then $\phi$ can be determined from Herschel's naked-eye limiting magnitude. An indication of this is that he found Uranus (5.9 mag at its faintest) a near-threshold object \citep[v.1, p.106]{herschel}, but more precise is his remark on double star H I 69 (CCDM J07057+5245AB) which he discovered in November 1782: `in a very clear evening it may just be seen with the naked eye' \citep[v.1, p.333]{herschel}. This system, which Herschel would have been able to view almost exactly at the zenith, has integrated magnitude 6.12 and colour index 0.1. The average colour index of objects in the NGC is 0.85 \citep{steinickeni}, for which the corresponding limit would be 5.92 (from Eq. \ref{starcorr}); but Herschel may have been able to see objects slightly fainter than H I 69. As an approximation his limit will be taken as 6.0 mag (with respect to assumed colour index 0.85).

It will be noted that nearly twenty years elapsed between Herschel's naked-eye star observation (aged 44) and his measurement of his eye pupil (aged 63), and it is entirely possible that both figures would have changed over that period. One could also question the assumption $\phi = \sqrt{2}F$ since the factors are not for equivalent search procedures: Herschel's telescopic search was for objects not previously known, whereas his naked-eye observation was of a star whose position he knew in advance. Moreover, the front-view mode would have introduced aberration because the mirror was viewed at an angle to the optical axis, and more generally his telescope cannot have been optically perfect. Nevertheless the stated figures will be adopted for calculation purposes.

From Eq. \ref{deltai} with the assumed sky brightness we find $F = 2.36$. From Eq. \ref{m} this implies that Herschel's limit for stars seen with the telescope at sweeping power $M = 157$ would have been 15.66 mag in front-view mode and 0.24 mag poorer in Newtonian mode (assuming 67 per cent reflectance for the secondary, and without correction for the central obstruction whose size is not recorded). This is consistent with the magnitude limit of his catalogue: out of roughly 2,500 nebulae Herschel discovered, only 7 are 15.0 mag or fainter \citep{steinickengc}, down to a minimum of 15.5 mag for NGC 2843 and NGC 4879, the latter being a mis-identified star. For verifying objects Herschel sometimes used $M=240$ \citep[v.1, p.268]{herschel}, with a predicted front-view limit 16.08 mag. The cut-off for $M \geq 328$ would have been 16.39 mag. 

Inserting the parameter values into Eq. \ref{scurvetel} gives Herschel's telescopic threshold curve at the sweeping power he used:
\begin{equation}
\mu_{157} = 23.18 - 4.167\mathrm{log}(0.468\alpha^{-0.6} + 1).
\label{157}
\end{equation}

\begin{figure}
\includegraphics[width=84mm]{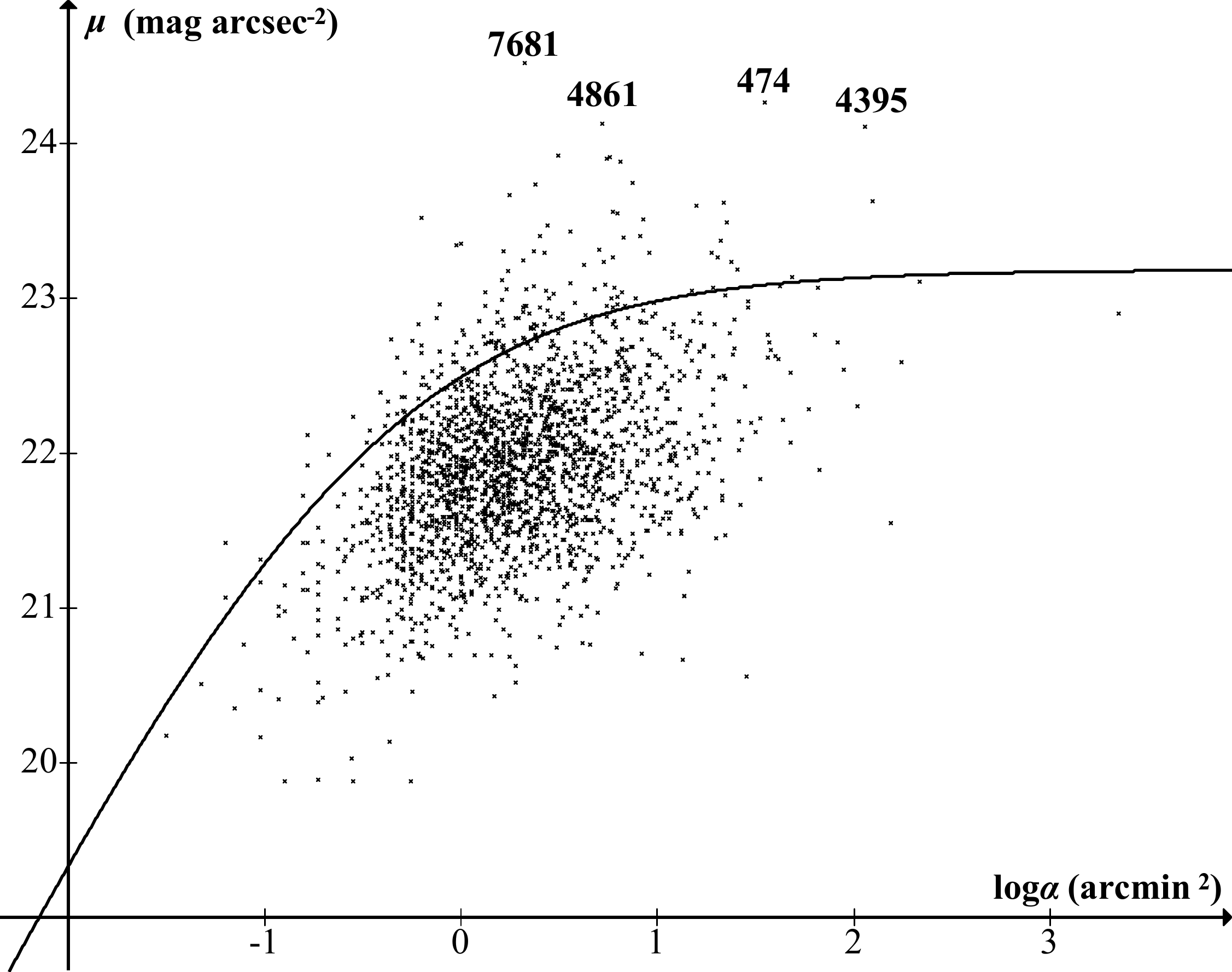}
\caption{Objects found by William Herschel, plotted by surface brightness $\mu$ versus area $\alpha$. 91.7 per cent lie below his predicted threshold curve (Eq. \ref{157}); some extreme outliers (labelled by NGC number) are discussed in the text.}
\label{herschel}
\end{figure}

It is difficult to verify this precisely since astrophysical data are skewed by isophotal limit, and Herschel needed only to see the bright centre of an object in order to detect it. In view of these limitations, Fig \ref{herschel} plots all 2136 Herschel objects for which magnitude and area data are given in \citet{steinickeni} and \citet{steinickengc}, without correction for colour index or zenith angle. It can be seen that the great majority (91.7 per cent) lie below the predicted threshold. Only 3.2 per cent are more than 0.25 mag arcsec\textsuperscript{-2} above it.

Some extreme outliers are marked on Fig. \ref{herschel} by their NGC numbers; in all cases Herschel did not see the complete object. Herschel estimated NGC 4395 (H V 29) as `10 arcmin long, 8 or 9 arcmin broad' \citep[v.1, p.359]{herschel}, which is only about 60 per cent of its actual area (three of its H\,{\sc ii} regions became designated as separate nebulae in the NGC). He saw NGC 4861 (H IV 30) as `two stars, distance 3 arcmin, connected with a very faint narrow nebulosity' \citep[v.1, p.356]{herschel}, but the galaxy is approximately 40 per cent longer. He described NGC 7681 (H II 242) as `small' \citep[v.1, p.275]{herschel} and NGC 474 (H III 251) as `very small' \citep[v.1, p.284]{herschel}.

\begin{figure}
\includegraphics[width=84mm]{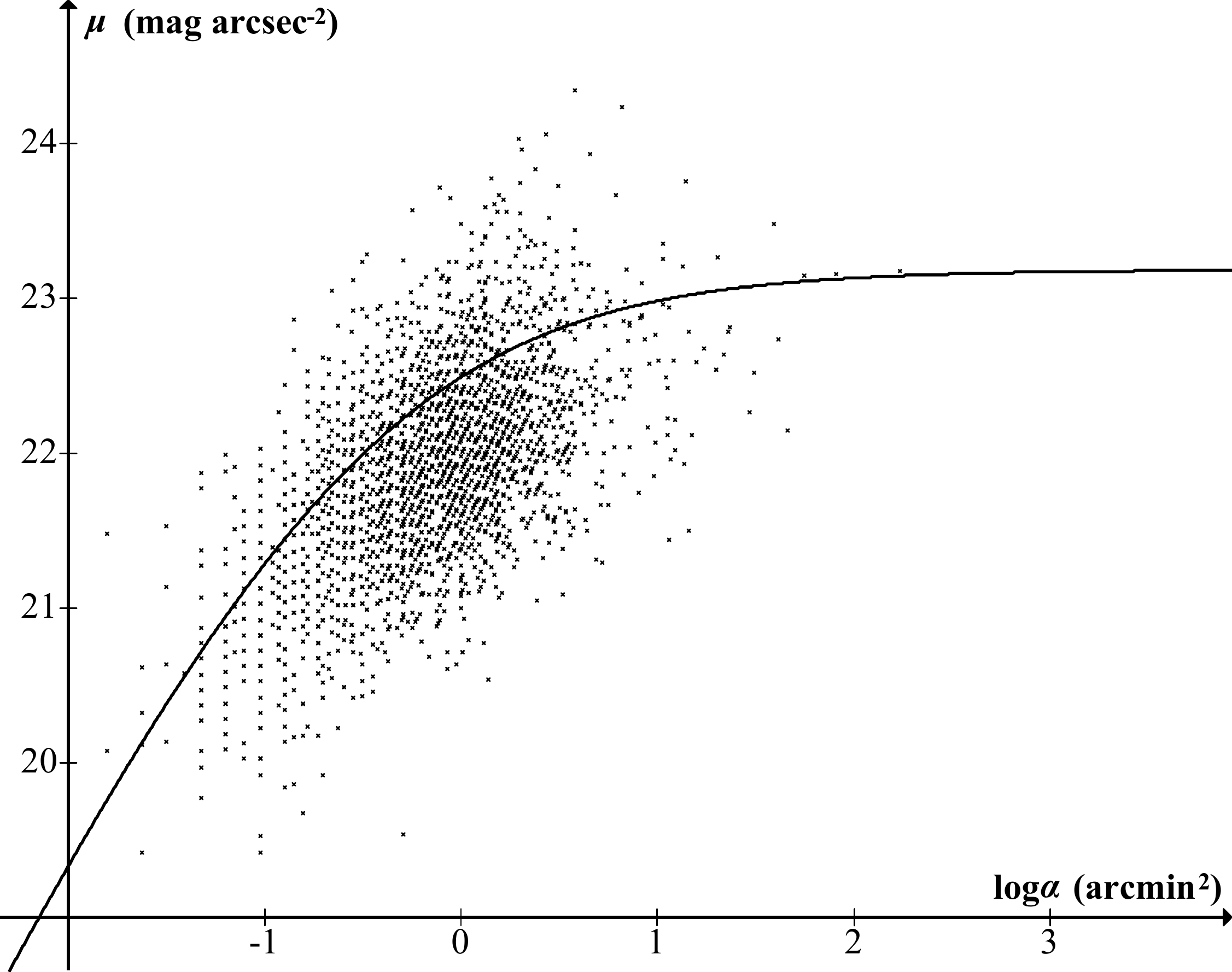}
\caption{NGC objects missed by William Herschel. These are generally smaller, and nearer to his threshold, than those in Fig. \ref{herschel}. 90.6 per cent of missed objects below the curve are smaller than the average size of objects he detected.}
\label{missedherschel}
\end{figure}

Fig. \ref{missedherschel} shows NGC objects which Herschel failed to discover (3431 objects with declination higher than -33 degrees, unknown to Herschel, for which magnitude and area data are available in \citet{steinickeni} and \citet{steinickengc}, using Epoch 2000.0 co-ordinates and without atmospheric or photometric correction). Herschel did not sweep the entire sky above his horizon \citep[p.34]{steinicke}, and this can account for some of the omissions. Others could have been missed because of low declination, crowded search fields, proximity to glare sources (bright stars), limited search time or human error. In general, however, it can be seen that the missed objects are smaller and nearer threshold than the discovered ones. This can be quantified using `visibility level', defined as the ratio of an object's contrast to the threshold level \citep{adrian}, for which the corresponding astronomical quantity is the object's distance ($L$) below the curve. For the objects in Fig. \ref{herschel} mean $L$ is 0.69, and mean $\mathrm{log}\alpha$ is 0.28, whereas for the missed objects mean $L$ is 0.35, mean $\mathrm{log}\alpha$ is -0.21. Missed objects with $\alpha$ and $L$ larger than the mean values for discovered ones, and which therefore should have been easy targets for Herschel, amount to only 3.6 per cent of those he did not see, illustrating the thoroughness of his search.

\subsection{Further applications}

\begin{figure}
\includegraphics[width=84mm]{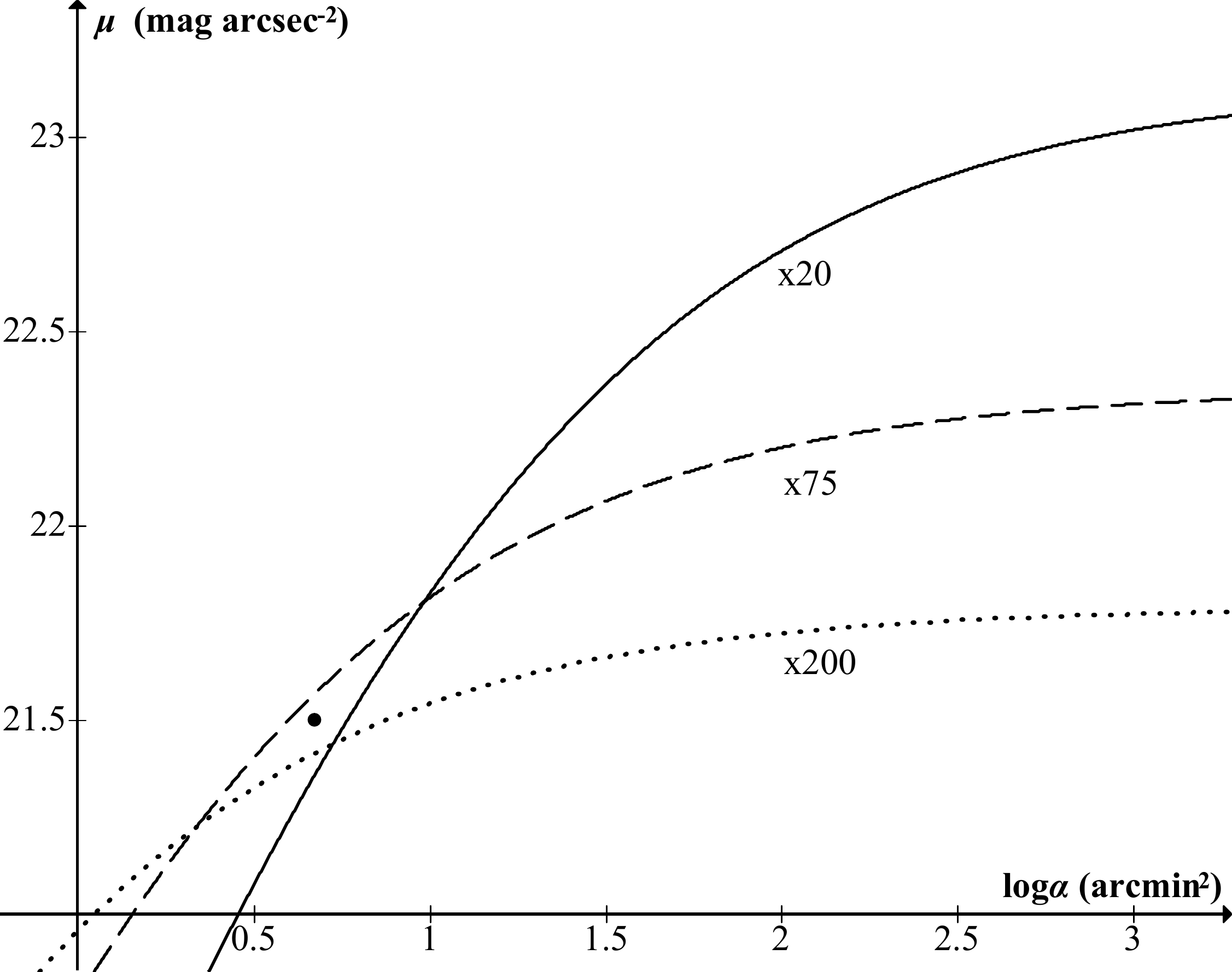}
\caption{Threshold curves for a 100mm telescope at various magnifications (with $\mu_\mathrm{sky}=21.4$ mag arcsec\textsuperscript{-2}, $p = 6$ mm, $F_{\mathrm{t}} = 1.04$, $\phi = 3.55$). The data point is for an object predicted to be visible at $\times 75$ but invisible at the lower and higher powers.}
\label{clark}
\end{figure}

There has been interest among visual astronomers in the concept of `optimum magnification' (\citet{lewis}, \citet{clark}, \citet{garstang1999}, \citet{clark2014}). The contrast of an extended object seen in a telescope is independent of magnification, but the threshold is dependent on image size and background, both of which change with magnification. Hence an object may be invisible at low or high power but visible in some intermediate range. This is represented in Fig. \ref{clark} which shows threshold curves for a 100mm telescope at magnifications 20, 75 and 200 (with parameters chosen for convenience of illustration), and a single data point representing a hypothetical non-stellar object. Raising power shifts the Ricco asymptote to the left (increasing the point-source limit) but lowers the horizontal asymptote (decreasing the surface-brightness limit). The object is predicted to be visible at magnification 75 but not at the lower and higher powers. The model could be used to obtain optimum magnifications for actual objects, but such predictions are of limited value, both because of the lack of appropriate data at the correct isophotal limit, and (most importantly) because targets are in general not uniform. It is however interesting to note the finding of \citet{leibowitz} that at low light levels visual acuity is greatest for a pupil size of approximately 3mm. This corresponds to the exit pupil chosen by William Herschel for his nebula sweeps, which he presumably arrived at using trial and error. The same optimum exit pupil was found independently by \citet{langley}.

\begin{figure}
\includegraphics[width=84mm]{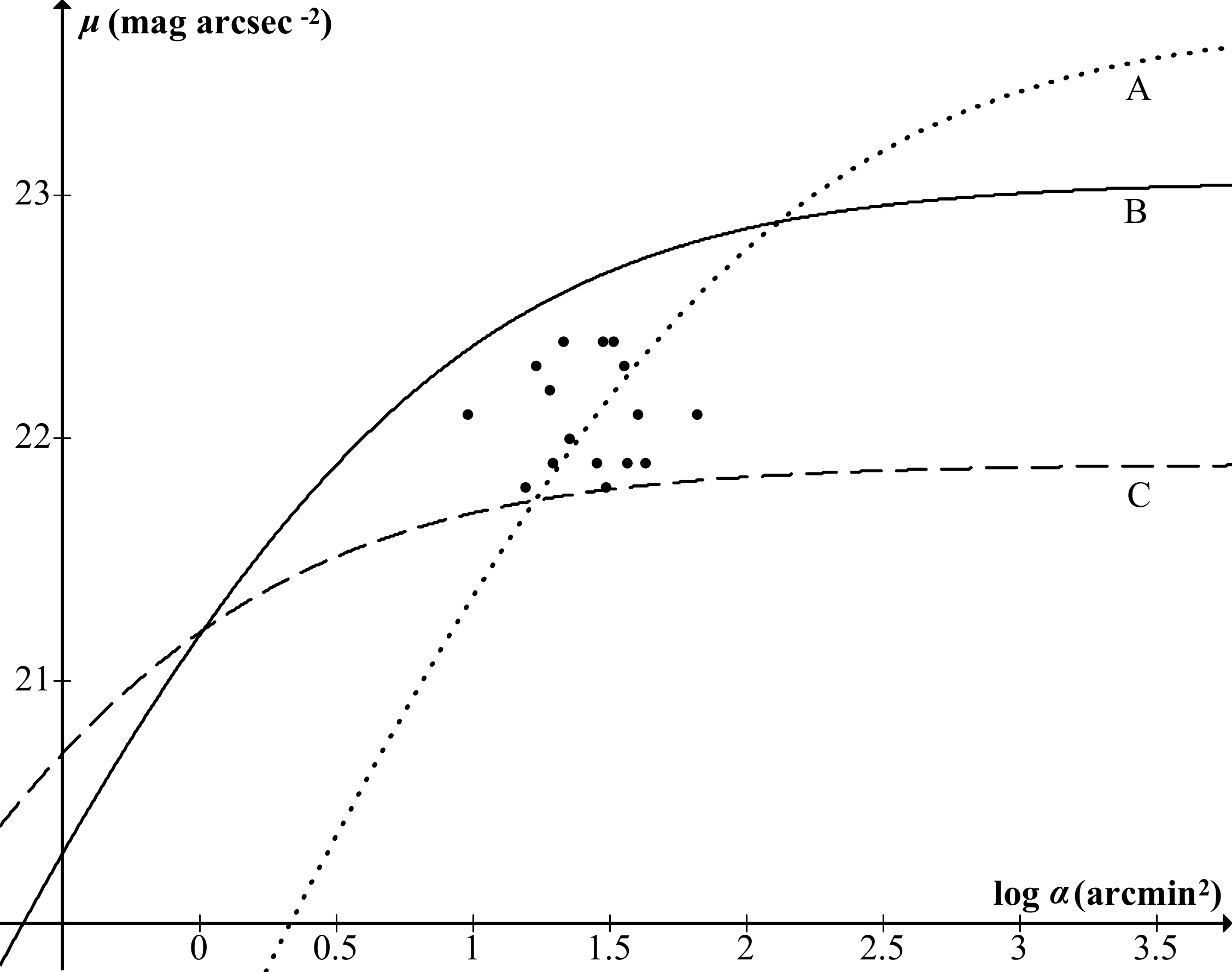}
\caption{Threshold curves for various viewing situations. Curve A is for $10 \times 50$ binoculars at a dark site ($\mu_\mathrm{sky} = 21.5$ mag arcsec\textsuperscript{-2}), curve B is a 6-inch refractor with exit pupil 3mm at the same site, and curve C is a 16-inch reflector with the same exit pupil at a light-polluted site ($\mu_\mathrm{sky} = 20$ mag arcsec\textsuperscript{-2}). Data points are for Messier galaxies in the Virgo Cluster.}
\label{comparo}
\end{figure}

It is interesting to make general comparative predictions of instrument performance. Fig. \ref{comparo} shows threshold curves for a single user ($p=7$ mm, $F=2$) at two sites, one light polluted ($\mu_\mathrm{sky} = 20$ mag arcsec\textsuperscript{-2}, naked-eye limit 5.5 mag), the other dark ($\mu_\mathrm{sky} = 21.5$ mag arcsec\textsuperscript{-2}, 6.0 mag). The instruments are $10 \times 50$ binoculars and a 6-inch refractor ($D = 150$ mm) at the dark site, and a 16-inch reflector with 25 per cent central obstruction ($D=393$ mm) at the light-polluted one, with assumed transmittances 85, 95 and 75 per cent respectively. Both telescopes have exit pupil 3mm. Data are also plotted for the 16 Messier galaxies in the Virgo Cluster \citep{steinickeni}, subject to the usual caveats regarding isophotal limit and non-uniformity, but providing a reasonably homogeneous sample for illustrative purposes. It can be seen that for any target larger than 1 arcmin\textsuperscript{2} the 16-inch is outperformed by the smaller telescope at the darker site: light pollution renders it ineffective for viewing galaxies. Binoculars outperform the 6-inch for very large, low surface-brightness objects; however the 6-inch will show numerous smaller targets. Since the effect of varying the field factor $F$ is to move all the curves up or down equally, this qualitative result will remain the same for individuals whose naked-eye limit is higher or lower than the chosen figure.

\section{Conclusions}

A new way has been presented for modelling achromatic threshold visibility data such as that of \citet{blackwell} or \citet{kth}, and has been shown to represent the laboratory data more accurately than previous models. For applications at low light levels a photometric correction is needed; this has been calculated for scotopic vision, applicable to astronomical observations within approximately one magnitude of threshold at sites without excessive light pollution. The new point-source model matches the data of \citet{bowen} more accurately than previous atempts by \citet{schaefer} and \citet{garstang2000}. The model for extended targets offers new insights into the visibility of `deep sky objects' such as galaxies, and has been shown to be consistent with the observations of William Herschel.

The basic relation of the model, shown in Figures \ref{rbtable8} and \ref{cinf}, was purely empirical, but one might question whether it has a physiological basis. Eq. \ref{rscot} implies $C \approx (r_{1}^2/A)B^{-1/2}$, the de Vries Rose law \citep{rose}, so the relation models the departure of the visual system from ideal quantum detection, which is presumed to arise at the stage of post-retinal processing. Because the visual system is quite different from a detector limited only by quantum efficiency, the results presented here are not expected to be applicable to CCD imaging. However, since a similar threshold curve is obtained for targets identified visually from photographic plates \citep{hubble}, one would expect applicability there. Any line of constant magnitude brighter than the Ricco asymptote will intersect the threshold curve, demarcating zones that are visible or otherwise. Hence the completeness of any magnitude-limited sample is dictated by the shape of the curve together with the luminosity function for the targets in question. This should apply for example to the catalogue of \citet{shapleyames}, for which an empirical completeness function was found by \citet{sandage}.

The photopic model of Section 2 would be applicable to daylight phenomena such as sunspots, though visibility of objects against the blue sky would require proper incorporation of chromaticity. This would also be the case for mesopic applications, e.g. observation at heavily light polluted sites where colour can be perceived. Successful modelling of situations such as these would require new experimental data sets, other than the achromatic ones used here. At light polluted sites where scotopic vision is possible (i.e. darker than about 19 mag arcsec\textsuperscript{-2}), the sky spectrum will have anthropogenic contributions for which a photometric correction is necessary, as shown in Section 1.3.

Photometers are available which measure both photopic and scotopic luminance. This would be useful in light pollution studies since S/P ratio varies with type of lighting: a moonlit country sky and a moonless suburban one polluted by fluorescent lighting could both give the same reading on a Sky Quality Meter (e.g. 20 $\mathrm{mag_{SQ} arcsec^{-2}}$, or `SQ 20'), but the country sky would be darker to scotopic vision since the S/P ratio of fluorescent lighting is higher than that of moonlight. If the meter were fitted with a removable scotopic filter and suitably calibrated then S/P ratios could be found and quoted in addition to photopic luminance.

The International Dark-Sky Association \citep{idsa} currently recognizes three classes of dark sky: Bronze (SQ $20.00 \-- 20.99$), Silver (SQ $21.00 \-- 21.74$) and Gold (SQ $\geq$ 21.75). A reading greater than 22 is `unlikely to be recorded' \citep{unihedronfaq}. The suggested limiting magnitudes (based on the Bortle Scale) are Bronze: $5.0 \-- 5.9$, Silver: $6.0 \-- 6.7$, Gold: $\geq$ 6.8. It is questionable whether 20 mag arcsec\textsuperscript{-2} can be considered dark, given the results shown in Fig. \ref{comparo}. Also it has been argued here that a definition of naked-eye limiting magnitude based on momentary glimpses is overly susceptible to scintillation, which is a local and variable effect. Consequently it has been suggested that currently recommended magnitude limits may be excessive, compared with limits that would be obtained for targets visible for an extended period. It has also been shown that the recommendations of the Bortle Scale with regard to the visibility of M33 are contradicted by the present model. Since the scale appears to be based on subjective judgment rather than rigorous data, its reliability appears questionable.

A practical definition of a dark sky would be one in which the Milky Way is capable of being seen. The non-uniformity of the Milky Way makes this problematic to model, and even if a particular region were chosen as standard, there remains the problem that existing luminance measurements are based on a surface brightness limit fainter than that of the eye, and are usually filtered to remove bright stars, whereas unresolved stars just beyond the visual limit may contribute a significant proportion of the light detectable by eye. An equivalent limiting magnitude could be found empirically: observers would view the sky through a variable filter, adjusting it until a chosen portion of the Milky Way was considered just visible, and they would also note the faintest stars visible at this setting. 

\citet{bigourdan} noted that the summer Milky Way became visible from Paris Observatory when the Sun reached 13 degrees below horizon. The corresponding sky brightness is dependent on local conditions but would have been approximately $20.2 \-- 20.3$ mag arcsec\textsuperscript{-2} \citep{patat}. Bigourdan further noted that with the Sun 15 degrees below horizon he was able to see faint NGC objects, while an angle of 16 degrees was sufficient for the faintest to become visible. That would indicate approximate values of 21.3 and 21.5 mag arcsec\textsuperscript{-2}. That suggests a three-tier dark-sky classification with proportionally diminishing SQ bands $20.25 \-- 21.24$ (`grey'), $21.25 \-- 21.74$ (`black') and $21.75-22.00$ (`pristine'). Recalling from Section 1.3 the \citet{cie2010} suggested limit for scotopic vision, one could add a `bright' band SQ $18.25 \-- 20.24$, and a `white' band SQ $< 18.25$, in which scotopic vision would not be achievable for naked-eye sky observation.

\begin{table}
\caption{Magnitude penalty $pen = m_{22} - m_0$ and surface brightness supplement $sup = \mu_\infty - m_0$ with respect to ideal conditions ($\mu_\mathrm{sky} = 22$ mag arcsec\textsuperscript{-2}), with approximate sky-quality banding.}
\label{limits}
\begin{tabular}{lcccccc}
&Pristine & . & Black &  .  & Grey & .\\
$\mu_\mathrm{sky}$ & 22.00 & 21.75 & 21.50 & 21.25 & 21.00 & 20.75\\
\hline
$pen$ & 0.00 & 0.10 & 0.20 & 0.30 & 0.40 & 0.49\\
$sup$ & 18.06 & 17.98 & 17.90 & 17.82 & 17.74 & 17.66\\
\end{tabular}

\end{table}
\begin{table}

\begin{tabular}{lcccccc}
& Grey & . & Bright & .  & . & . \\
$\mu_\mathrm{sky}$  & 20.50 & 20.25 & 20.00 & 19.75 & 19.50 & 19.25\\
\hline
$pen$  & 0.59 & 0.68 & 0.77 & 0.85 & 0.93 & 1.01\\
$sup$  & 17.58 & 17.49 & 17.40 & 17.32 & 17.22 & 17.13\\
\end{tabular}
\end{table}

The threshold curve is relative rather than absolute, and it has been shown that for a given observing situation there is an overall field factor $F$ which can often be eliminated from calculations. Suppose that for a given individual under particular observing conditions at an ideal site ($\mu_\mathrm{sky} = 22$ mag arcsec\textsuperscript{-2}) the naked-eye limit is $m_{22}$, and that at a site with greater sky brightness but otherwise equivalent observing conditions the same person's limit is $m_0$. Then the `penalty' $m_{22} - m_0$ (calculated using Eq. \ref{deltai}) is independent of $F$, and hence of observer. The difference $\mu_\infty - m_0$ (from Eq. \ref{sblim}) is likewise independent of $F$, i.e. the limiting surface brightness can be expressed as a `supplement' to be added to an individual's point-source limit for a given site. This is shown in Table \ref{limits}, together with corresponding banding. For example, a person whose limit is 6.0 mag at a `grey' site with $\mu_\mathrm{sky} = 21$ mag arcsec\textsuperscript{-2} (which would correspond to $F$ = 1.74) is predicted to have limiting surface brightness 23.74 mag arcsec\textsuperscript{-2} at that site. At a `pristine' site with $\mu_\mathrm{sky} = 21.75$ mag arcsec\textsuperscript{-2} that same person is predicted to have limits 6.3 mag and 24.28 mag arcsec\textsuperscript{-2}. Good linear approximations for naked-eye stellar limits (error $<0.05$ mag) are:

\begin{equation}
m_0 = 0.27\mu_\mathrm{sky}+0.8-2.5\mathrm{log}F  \\ (18 \leq \mu_\mathrm{sky} \leq 20),
\end{equation}

\begin{equation}
m_0 = 0.383\mu_\mathrm{sky}-1.44-2.5\mathrm{log}F \\ (19.5 \leq \mu_\mathrm{sky} \leq 22).
\end{equation}

\section*{acknowledgements}
This work was begun while the author was Visiting Fellow at Durham Institute of Advanced Study. The author warmly thanks Martin Banks (Berkeley) and Gordon Love (Durham) for invaluable conversations during that initial phase of the project. The author also thanks the referee, Dan Duriscoe, for useful comments and suggestions.

\end{document}